\documentclass[superscriptaddress, aps,prd, amsmath,amssymb,showpacs,showkeys, onecolumn]{revtex4-2}
\usepackage[dvips]{graphicx,color}
\usepackage{subfig}
\usepackage{times}
\usepackage{xcolor}
\usepackage[%
  colorlinks=true,
  urlcolor=blue,
  linkcolor=red,
  citecolor=blue
]{hyperref}
\usepackage{orcidlink}

\begin{document}
\title{Quasinormal modes and greybody factors of symmergent black hole}

\author{Dhruba Jyoti Gogoi \orcidlink{0000-0002-4776-8506}}
\email[Email: ]{moloydhruba@yahoo.in}
\affiliation{Department of Physics, Dibrugarh University,
Dibrugarh 786004, Assam, India.}

\author{Ali \"Ovg\"un \orcidlink{0000-0002-9889-342X}}
\email[Email: ]{ali.ovgun@emu.edu.tr}
\affiliation{Physics Department, Eastern Mediterranean University, Famagusta, 99628 
North Cyprus via Mersin 10, Turkey.}

\author{Durmu\c{s}~Demir
\orcidlink{0000-0002-6289-9635}}
\email[Email: ]{durmus.demir@sabanciuniv.edu}
\affiliation{Faculty of Engineering and Natural Sciences, Sabanc{\i} University, 34956 Tuzla, \.{I}stanbul, Turkey}

\begin{abstract}
Symmergent gravity is an emergent gravity framework in which gravity emerges guided by gauge invariance, accompanied by new particles, and reconciled with quantum fields. In this paper, we perform a detailed study of the quasinormal modes and greybody factors of the black holes in symmergent gravity. Its relevant parameters  are the quadratic curvature term $c_{\rm O}$ and the vacuum energy parameter $\alpha$. 
In our analyses, effects of the both parameters are investigated. Our findings suggest that, in both positive and negative direction, large $|c_{\rm O}|$ values of the parameter on the quasinormal modes parallel the Schwarzschild black hole. Moreover, the quasinormal model spectrum is found to be sensitive to the symmergent parameter $\alpha$.  We contrast the asymptotic iteration and  WKB methods in regard to their predictions for the quasinormal frequencies, and find that they differ (agree) slightly at small (large) multipole moments. 
We analyze time-domain profiles of the perturbations, and determine the greybody factor of the symmergent black hole in the WKB regime. The symmergent parameter $\alpha$ and the quadratic curvature term $c_{\rm O}$ are shown to impact the greybody factors significantly. We provide also 
rigorous limits on greybody factors for scalar perturbations, and reaffirm the impact of model parameters.

\end{abstract}
\date{\today}


\maketitle

%\tableofcontents
\section{Introduction}

It is convenient to start by briefly discussing the whys, hows and whats of the symmergent gravity. For that purpose, let us consider a general quantum field theory (QFT) valid up to a UV cutoff scale $\Lambda_\wp$. It could be general enough new fields in excess of the known particles in the Standard Model of elementary particles (SM). At the loop level, this QFT develops power-law sensitivities to the UV cutoff: There arise ${\mathcal{O}}\left(\Lambda_\wp^2\right)$ shifts in scalar masses, gauge bosons acquire ${\mathcal{O}}\left(\Lambda_\wp^2\right)$ masses, and vacuum energy shift by ${\mathcal{O}}\left(\Lambda_\wp^4\right)$ contributions \cite{weinberg0}. These power-law UV sensitivities render the effective QFT  unphysical for describing physics at the QFT scale as is revealed in especially the detached regularization and renormalization scheme of \cite{demir0a}.  Quadratic corrections to the scalar and gauge boson masses give rise, respectively, to the gauge hierarchy and charge conservation problems \cite{demir3,demir2}. Symmergent gravity (short for gauge symmetry restoring gravity) is a novel approach in which these problems are solved in analogy with the Higgs mechanism \cite{demir0}. Indeed, masses of vector bosons are promoted to Higgs field to revive gauge symmetries as both the vector masses and the Higgs field respect Poincare symmetries. In parallel with this, the loop-induced gauge boson masses are promoted to affine curvature to revive gauge symmetries as both the UV cutoff and curvature break the Poincare symmetries. This metric-affine gravity leads dynamically to the GR after integrating out the affine connection (effectively giving a holographic UV completion of the QFT). Gravitation emerges from within the effective theory and lives in concordant with the quantum loop effects. Induction of the gravitational constant at the right scale necessitates introduction of new massive fields, with the highly interesting property that these new fields do not have to interact with the known particles in the SM \cite{demir0,demir1,demir2,demir3}. In summary, gravity emerges in a way guided by gauge invariance, reconciled with quantum fields, and accompanied by new particles  in the symmergent gravity framework. It possesses signatures that can be probed at collider experiments like the LHC \cite{lhc1}, celestial bodies like the black holes \cite{javlonla,reggiealiben,irfanaliben,cqgirfan,bh4}, cosmic events like inflation \cite{inf1}, and elusive matter like the dark matter \cite{dark-sector}. In the present work, our goal is to probe symmergent gravity effects via the quasinormal modes and greybody factors of black holes. To this end, in Sec. II below, we give a brief yet comprehensive discussion of the symmergent gravity and determine its black hole solutions before starting the phenomenological analyses in Sec. III and onward.  

%This symmetry-restoring emergent gravity, symmergent gravity in brief, predicts new particles beyond the known ones and has implications therefore for astrophysical, cosmological and collider phenomena. 

%Symmergent gravity is a whole new framework in which gravity emerges in a way guided by gauge invariance \cite{demir2,demir3}, reconciled with quantum fields \cite{demir1}, and accompanied by new particles \cite{demir1,demir2}. It can be searched at collider experiments like LHC, astrophysical objects like black holes, cosmological phenomena like inflation, and elusive species like dark matter. Symmergent gravity is an emergent gravity theory in which gravity emerges along with a plethora of new particles. In this approach, affine curvature is introduced to revive gauge symmetries broken explicitly by the UV cutoff just as the Higgs field is introduced to revive the gauge symmetry in massive vector fields  \cite{demir0}.

Ubiquitous vibrational patterns pervade our surroundings, from musical notes to expansive research fields such as seismology, asteroseismology, molecular spectroscopy, atmospheric studies, and structural engineering. Each of these fields focuses on understanding the structure and material composition through the lens of these unique vibrational patterns, akin to the proverbial "hearing the shape of a drum" \cite{Kac:1966xd}.  The focus of this paper is the distinctive oscillations of symmergent black holes, known as quasinormal modes (QNMs) \cite{Andersson:1992scr,Andersson:1994rm,Andersson:1995vi,Andersson:1996xw,Maggio:2019zyv,Konoplya:2022zym,Konoplya:2018yrp,Berti:2005ys}. The discovery of gravitational waves (GWs) ushers in a new era of gravitational physics research and heralds the onset of GW astronomy \cite{
LIGOScientific:2016aoc}-\cite{Yang:2019vni}. While all orbiting celestial bodies generate GWs, only compact, rapidly moving objects can produce signals strong enough for current detectors, making black hole binaries ideal for GW detection. A black hole collision consists of three stages: inspiral, where black holes orbit each other, closing in due to energy loss through GWs; merger, the actual collision; and ringdown, where the merged black hole attains its equilibrium form, often a Kerr black hole. The resulting GW signal carries unique signatures, disclosing the black hole's properties. In a disturbed system, energy dissipation-related vibrations are termed (QNMs), which define the ringdown phase of black hole coalescence \cite{Cardoso:2016rao}-\cite{Cardoso:2017cqb}. The quest to find QNM frequencies originated with Regge, Wheeler, and Zerilli's work, viewing the problem as a non self-adjoint boundary problem. This approach results in a complex spectrum, with the real part representing oscillation frequency, and the imaginary part denoting wave damping \cite{Pantig:2022gih}-\cite{Chabab:2017knz}. In recent years, Einstein's General Relativity (GR) theory has seen substantial empirical confirmations, yet many believe it is not the final theory of gravity. Certain scenarios such as the early universe, black hole interiors, and final stages of black hole evaporation demand a quantum theory explanation where GR falls short, highlighting the need to reconcile gravity with quantum mechanics \cite{Lambiase:2023fbd}.

We aim to explore the underlying theory of symmergent gravity, the insights they provide about these cosmic entities, and their intersections with other physics domains. Black hole environments, unlike most large-scale physical systems, inherently exhibit dissipation due to an event horizon, thus ruling out a typical normal-mode analysis \cite{Vishveshwara, Press, Chandrasekhar_qnms}. As such, these systems lack time-symmetry and pose a non-Hermitian boundary value problem. Generally, QNMs possess complex frequencies, where the imaginary component represents the decay timescale of the disturbance. The associated eigenfunctions often defy normalization, and they rarely constitute a complete set. Given the prevalence of dissipation in real-world physical systems, it's unsurprising that QNMs have extensive applicability, including in the handling of many dissipative systems, such as atmospheric phenomena and leaky resonant cavities. Recent studies show that QNMs can play a significant role in distinguishing different modified theories of gravity \cite{qnm_bumblebee, gogoi3, Graca, Zhang2, Liang2018, Hu, hemawati2022, gogoi4, Ovgun:2017dvs, Rincon:2018sgd, Panotopoulos:2019qjk, Panotopoulos:2020mii, Rincon:2021gwd,Gonzalez:2021vwp,Zhidenko:2003wq,Zhidenko:2005mv,Ovgun:2019yor,gogoi_wormhole, 37,40-17, Sekhmani:2023ict, fQ,Guo:2020blq,Kuang:2017cgt}. In the near,  future, Laser Interferometer Space Antenna (LISA) can help us to put a stringent constraint on different modified theories of gravity and black hole spacetimes \cite{LISA:2017pwj}.

Greybody factors represent the transmission probabilities of Hawking radiation, a form of emission from black holes resulting from their gravitational potential. In classical terms, a black hole can only absorb particles, not emit them \cite{Hawking:1975vcx,Singleton:2011vh,Akhmedova:2008dz}. However, through quantum effects, a black hole can generate and emit particles from its event horizon, manifested as thermal radiation known as Hawking radiation. This radiation interacts with the black hole's self-created gravitational potential, leading to the reflection and transmission of the radiation. Consequently, the spectrum observed by a distant observer deviates from the typical blackbody spectrum. A black hole greybody factor quantifies this departure from pure blackbody radiation \cite{Maldacena:1996ix,Cvetic:1997uw,Fernando:2004ay}. Various methods exist to calculate the greybody factor, one of which involves a strict bound for reflection and transmission coefficients for one-dimensional potential scattering, a formulation applicable to a black hole's greybody factor \cite{Okyay2022,Ovgun188,Pantig:2022gih,Yang:2022xxh,Yang:2022ifo,Panotopoulos:2018pvu,Panotopoulos:2016wuu,Rincon:2018ktz,Ahmed:2016lou,Javed:2021ymu,Javed:2022kzf,Mangut:2023oou,Al-Badawi:2022aby}.

The paper is organized as follows: In Sec. II, we provide a brief and comprehensive discussion of the symmergent gravity, and give its black hole solutions. In Sec. III, we give a comparative discussion of the scalar and vector perturbations, and derive the corresponding quasinormal modes (QNMs). In Sec. 
IV, we investigate the time evolution profiles of the perturbations in Sec. III.  In Sec. V, we study the greybody factors using the WKB approach.  Finally, in Sec. VI, we conclude the paper by discussing the results and giving the future prospects in Sec. VI. 

\section{Symmergent Gravity and Its Black Hole Solution} \label{sec2}

Here in this section we give a brief yet comprehensive discussion of the symmergent gravity (with particular emphasis on its parameter space). Let us start with a general QFT, which can contain new fields compared to the known ones in the SM.  QFTs live inherently in flat spacetime since only in there that a Poincar\'e-invariant vacuum state becomes possible \cite{incompatible,wald}. They are defined by an invariant action and a UV cutoff $\Lambda_\wp$. Incorporation of gravity into QFTs in flat spacetime can be accomplished either by taking the QFT to curved spacetime  (hampered by Poincar\'e breaking in curved spacetime \cite{dyson,wald}) or by quantizing the gravity itself (seems absent presently \cite{thooft}). These two obstacles leave no room other than the emergent gravity approach \cite{sakharov,visser,verlinde}.

In general, particle masses conserve the Poincar\'e symmetry as they are Casimir invariants of the Poincar\'e group. But there can exist Poincare-breaking scales. The UV cutoff $\Lambda_\wp$ is one such scale. And, according to the findings of \cite{fn2}, loss of Poincar\'e invariance could be interpreted as the emergence of gravity from within the QFT. This means that curvature can emerge from the Poincar\'e breaking sources in a QFT, and the hard momentum cutoff $\Lambda_\wp$ is the only Poincar\'e breaking source in a flat spacetime QFT. Indeed, the UV  cutoff $\Lambda_\wp$ \cite{cutoff} limits loop momenta $p_\mu$ as $-\Lambda_\wp^2 \leq \eta^{\mu\nu} p_\mu p_\nu \leq \Lambda_\wp^2$, and the loop corrections across this momentum interval change the QFT action $S[\eta,\phi,V]$ of scalars $S$ and gauge bosons $V_\mu$ as 
\begin{eqnarray}
 \delta S[\eta,\phi,V;\Lambda_\wp]  = \int d^4 x \sqrt{-\eta} \left\{-V_{\rm O} -c_{\rm O} \Lambda_\wp^4 - \sum_i c_{m_i} m_i^2 \Lambda_\wp^2 - c_S \Lambda_\wp^2 S^\dagger S + c_V \Lambda_\wp^2 V_{\mu} V^{\mu} \right\}
 \label{action-0}
\end{eqnarray}
in $(+,-,-,-)$ metric signature with the flat metric  $\eta_{\mu\nu}$, vacuum energy $V_{\rm O}$ (not power-law in $\Lambda_\wp$), masses $m_i$ of the QFT field $\psi_i$ (summing over all the fermions and bosons),  and the loop factors $c_{\rm O}$, $c_m$, $c_S$ and $c_V$ describing, respectively, the quartic vacuum energy correction, quadratic vacuum energy correction, quadratic scalar mass correction, and the loop-induced gauge boson mass \cite{gauge-break1,gauge-break2}. The gauge boson mass term $c_V \Lambda_\wp^2 V_{\mu} V^{\mu}$ reveals that the UV cutoff $\Lambda_\wp$ breaks gauge symmetries explicitly as because it is not the mass of a particle namely it is not a   Casimir invariant of the Poincar\'e group. \cite{demir2,demir3}.

Recalling induced gravity theories such as Sakharov's induced gravity \cite{sakharov,visser} proves useful for getting a sense of what to do with the power-law corrections in (\ref{action-0}). In induced gravity, the cutoff $\Lambda_\wp$ is readily identified with the gravitational scale but this leaves gauge symmetries explicitly broken with Planckian 
vacuum energy and Planckian scalar masses. The attempts made in \cite{demir2,demir3} was to improve the gauge sector in Sakharov's setup and this has led to a completely new structure \cite{demir0,demir1} -- gauge symmetry-restoring emergent gravity or, briefly, the symmergent gravity. In this aim, one first takes the effective QFT in (\ref{action-0}) (not the classical field theory) to curved spacetime of a metric $g_{\mu\nu}$. But, in an effective QFT couplings, masses and fields are all loop-corrected, and addition of any curvature term to make  $g_{\mu\nu}$ dynamical results in a manifestly inconsistent action.  Actually, it is not difficult to see that curvature can arise in the effective QFT in \eqref{action-0} only in the vector field (or tensor field) sector. The reason is that  the Ricci curvature $R_{\mu\nu}({}^g\Gamma)$, for instance, arises via the commutator $[\nabla_\lambda,\nabla_\mu]V^\lambda=R_{\mu\nu}({}^g\Gamma)V^\nu$ in which $\nabla_\mu$ is the covariant derivative with respect to the Levi-Civita connection ${}^g\Gamma^\lambda_{\mu\nu}$ of the curved metric $g_{\mu\nu}$. 
In view of this property, the effective QFT can be taken to curved spacetime as \cite{demir0,demir1,demir2}
\begin{eqnarray}
 S_{eff}[\eta,\phi,V,\psi ; \Lambda_\wp]
 \longrightarrow    S_{eff}[g,\phi,V,\psi ; \Lambda_\wp] - \int d^4 x \sqrt{-g} c_V V_{\mu} R^{\mu\nu}(g) V_\nu
 \label{map-eff-action}
\end{eqnarray}
in accordance with general covariance ($\eta_{\mu\nu}\rightarrow g_{\mu\nu}$ and $\partial_\mu \rightarrow \nabla_\mu$) \cite{covariance}. This transformation of the effective action implies 
\begin{eqnarray}
c_V \Lambda_\wp^2 \eta^{\mu\nu} V_{\mu} V_{\nu}
\longrightarrow c_V V_{\mu} \left( \Lambda_\wp^2 g^{\mu\nu} - R^{\mu\nu}(g)\right) V_\nu
\label{vector-mass-flat-2-curved}
\end{eqnarray} 
as the transformation to curved spacetime of the gauge-symmetry breaking terms in the power-law effective action in \eqref{action-0}.   

Having taken the flat spacetime effective QFT (whose power-law part is \eqref{action-0}) into curved spacetime of a metric $g_{\mu\nu}$, it is now time to analyze what to do with the UV cutoff $\Lambda_\wp$. To this end, let us first consider the mass term $M_V^2 V_\mu V^\mu$ for a massive vector field $V_\mu$. It is well known that gauge symmetry in the vector field $V_\mu$ can be restored by promoting $M_V^2$ to dynamical Higgs field $H$ with the replacement \cite{anderson,englert,higgs}
\begin{eqnarray}
M_V^2 V_\mu V^\mu\longrightarrow (V_\mu H)^\dagger (V^\mu H) \subset (D_\mu H)^\dagger (D^\mu H)
\label{higgs-map}
\end{eqnarray} 
in which the second equality concerns the Higgs kinetic term with the gauge-covariant derivative $D_\mu$.  This replacement is right in that both the mass $M_V$ and the Higgs field $H$ are Poincare-conserving quantities \cite{demir0}. Now, let's go back to the power-law effective action \eqref{action-0}). In analogy with the Higgs mechanism above, we may consider restoring the gauge symmetries by promoting  $\Lambda_\wp$ to a Higgs-like field. It is not, however,  as trivial as it seems. The reason is that, unlike the vector boson mass $M_V$,  the UV cutoff $\Lambda_\wp$ breaks the Poincare symmetry and the Higgs-like field in question must be a Poincare-breaking field. It has long been known that the Poincare-breaking field should be the spacetime curvature \cite{fn2}. But, since the loop-induced gauge boson mass term $c_V \Lambda_\wp^2 V_{\mu} V^{\mu}$ in \eqref{action-0}) exists in flat spacetime the curvature in question cannot be the curvature of the metric $g_{\mu\nu}$ as it vanishes in the flat limit \cite{demir2,demir3}. The resolution is to use the affine curvature which remains nonzero in flat spacetime and which dynamically approaches to the metrical curvature in curved spacetime \cite{demir0,demir1}. Then, in analogy with the promotion in (\ref{higgs-map}), the map \eqref{vector-mass-flat-2-curved} can be furthered as 
\begin{eqnarray}
c_V V_{\mu} \left( \Lambda_\wp^2 g^{\mu\nu} - R^{\mu\nu}(g)\right) V_\nu \longrightarrow c_V V_{\mu} \left( {\mathbb{R}}^{\mu\nu}(\Gamma) - R^{\mu\nu}(g)\right) V_\nu
\label{vector-map-affine}
\end{eqnarray} 
in which ${\mathbb{R}}^{\mu\nu}(\Gamma)$ is the Ricci curvature of the affine connection $\Gamma^{\lambda}_{\mu\nu}$ -- a general affine connection which is completely independent of the curved metric $g_{\mu\nu}$ and its Levi-Civita connection \cite{affine1,affine2,affine3}. This has the meaning that the UV cutoff $\Lambda_\wp$ is promoted to affine curvature as
\begin{eqnarray}
\Lambda_\wp^2 g^{\mu\nu} \rightarrow {\mathbb{R}}^{\mu\nu}(\Gamma)
\label{map}
\end{eqnarray}
in parallel with the promotion $M_V \rightarrow H$ of the vector boson mass $M_V$ in \eqref{higgs-map}. Under this map the power-law effective action in \eqref{action-0} takes the form 
\begin{eqnarray}
 \delta S[g,\phi,V, {\mathbb{R}}]  = \int d^4 x \sqrt{-\eta} \left\{-V_{\rm O}-\frac{c_{\rm O}}{16} {\mathbb{R}}^2(g) - \sum_i \frac{c_{m_i}}{4} m_i^2 {\mathbb{R}}(g) - \frac{c_S}{4} {\mathbb{R}}(g) S^\dagger S + c_V V_{\mu} \left({\mathbb{R}}^{\mu\nu}(\Gamma)-R^{\mu\nu}(g)\right)V_\nu \right\}
 \label{action-1}
\end{eqnarray}
in which ${\mathbb{R}}(g)\equiv g^{\mu\nu} {\mathbb{R}}_{\mu\nu}(\Gamma)$ is the scalar affine curvature \cite{demir1}. This metric-Palatini theory  contains  both the metrical curvature $R(g)$ and the affine curvature ${\mathbb{R}}(\Gamma)$ \cite{affine3,biz-beyhan}. From the third term, Newton's gravitational constant $G$ is read out to be 
\begin{eqnarray}
G^{-1}= 4 \pi \sum_i c_{m_i} m_i^2 \xrightarrow{\rm one\ loop} \frac{1}{8\pi} {\rm str}\!\left[{\mathcal{M}}^2 \right]
\label{MPl}
\end{eqnarray}
in which ${\mathcal{M}}^2$ is the mass-squared matrix of all the fields in the QFT spectrum. In the one-loop expression,  ${\rm str}[{\mathcal{M}}^2] = \sum_i (-1)^{2s_i} (2 s_i +1) {\rm tr}[{\mathcal{M}}^2]_{s_i}$ in which ${\rm tr}[\dots]$
is the usual trace (including the color degrees of freedom), $s_i$ is the spin of the QFT field ($s_i=0,1/2,\dots$), and $[{\mathcal{M}}^2]_{s_i}$ is the mass-squared matrix of the fields having that spin $s_i$. It is clear from the expression \eqref{MPl} that the known particles in the SM can generate Newton's constant neither in sign nor in size. It therefore is necessary to introduce new massive particles beyond SM spectrum. so that the supertrace in \eqref{MPl} leads to the correct gravitational scale. It is highly interesting that these new particles do not have to couple to the SM particles since the only constraint on them is the super-trace in (\ref{MPl}) \cite{demir0,demir1,demir2}. 

The metric-Palatini gravity in (\ref{action-1}) is not yet a proper gravitational theory. It is necessary to integrate out the affine connection $\Gamma^\lambda_{\mu\nu}$ to get the requisite metrical theory. This requires solving the motion equation of   $\Gamma^\lambda_{\mu\nu}$
\begin{eqnarray}
\label{gamma-eom}
{}^{\Gamma}\nabla_{\lambda}{{\mathbb{D}}}_{\mu\nu} = 0
\end{eqnarray}
in which ${}^{\Gamma}\nabla_{\lambda}$ is covariant derivative with respect to the affine connection $\Gamma^\lambda_{\mu\nu}$, ${\mathbb{D}}_{\mu\nu}= (g/{\rm Det}[{\mathbb{Q}}])^{1/6} {\mathbb{Q}}_{\mu\nu}$, and 
\begin{eqnarray}
\label{q-tensor}
{{\mathbb{Q}}}_{\mu\nu} = \left(\frac{1}{16\pi G} +   \frac{c_\phi}{4} \phi^\dagger\phi + \frac{{\overline{c}}_O}{8} g^{\alpha\beta} {\mathbb{R}}_{\alpha\beta}(\Gamma)\right) g_{\mu\nu} - c_{V} {\mbox{tr}}\left[V_{\mu}V_{\nu}\right]
\end{eqnarray}
is a bosonic field tensor, including the affine curvature ${\mathbb{R}}(\Gamma)$. 
The motion equation (\ref{gamma-eom}) implies that ${{\mathbb{D}}}_{\mu\nu}$ is covariantly-constant with respect to $\Gamma^\lambda_{\mu\nu}$, and this constancy leads to the exact  solution
\begin{eqnarray}
\Gamma^\lambda_{\mu\nu} &=& \frac{1}{2} \left({{ \mathbb{D}}}^{-1}\right)^{\lambda\rho} \left( \partial_\mu {{ \mathbb{D}}}_{\nu\rho} + \partial_\nu {{ \mathbb{D}}}_{\rho\mu} - \partial_\rho {{ \mathbb{D}}}_{\mu\nu}\right) = {}^g\Gamma^\lambda_{\mu\nu} + \frac{1}{2} ({{\mathbb{D}}}^{-1})^{\lambda\rho} \left( \nabla_\mu {{ \mathbb{D}}}_{\nu\rho} + \nabla_\nu {{ \mathbb{D}}}_{\rho\mu} - \nabla_\rho {{\mathbb{D}}}_{\mu\nu}\right)
\label{aC}
\end{eqnarray}
where ${}^g\Gamma^\lambda_{\mu\nu}$ is the Levi-Civita connection of $g_{\mu\nu}$. Given the smallness of gravitational constant in (\ref{MPl}), it is legitimate to make the expansions
\begin{eqnarray}
\Gamma^{\lambda}_{\mu\nu}&=&{}^{g}\Gamma^{\lambda}_{\mu\nu} + 8\pi G \left( \nabla_\mu {\overline{\mathbb D}}^\lambda_\nu + \nabla_\nu {\overline{\mathbb D}}^\lambda_\mu - \nabla^\lambda {\overline{\mathbb D}}_{\mu\nu}\right) + {\mathcal{O}}\left(G^2\right)
\label{expand-conn}
\end{eqnarray}
and
\begin{eqnarray}
{\mathbb{R}}_{\mu\nu}(\Gamma) &=& R_{\mu\nu}({}^{g}\Gamma) + 8\pi G\left(\nabla^{\alpha} \nabla_{\mu} {\overline{\mathbb D}}_{\alpha\nu} + \nabla^{\alpha} \nabla_{\nu} {\overline{\mathbb D}}_{\alpha\mu} - \Box {\overline{\mathbb D}}_{\mu\nu} - \nabla_{\mu} \nabla_{\nu} {\overline{\mathbb D}}_{\alpha}^{\alpha}\right)  +  {\mathcal{O}}\left(G^2\right)
\label{expand-curv}
\end{eqnarray}
so that both $\Gamma^{\lambda}_{\mu\nu}$ and ${\mathbb{R}}_{\mu\nu}(\Gamma)$ contain pure derivative terms  at  
the next-to-leading order through the reduced bosonic field tensor  ${\overline{\mathbb D}}_{\mu\nu}=
\left(\frac{c_\phi}{12} \phi^\dagger\phi + \frac{{\overline{c}}_O}{24} g^{\alpha\beta} {\mathbb{R}}_{\alpha\beta}(\Gamma)+\frac{c_V}{6} g^{\alpha\beta} {\mbox{tr}}\left[V_{\alpha}V_{\beta}\right]\right) g_{\mu\nu} - c_{V} {\mbox{tr}}\left[V_{\mu}V_{\nu}\right]$ \cite{demir1,demir2,demir3,bks}. The expansion in (\ref{expand-conn}) ensures that the affine connection $\Gamma^{\lambda}_{\mu\nu}$ is solved algebraically order by order in $G$  despite the fact that its motion equation (\ref{gamma-eom}) involves its own curvature ${\mathbb{R}}_{\mu\nu}(\Gamma)$ through ${\mathbb{D}}_{\mu\nu}$   \cite{affine1,affine2}. The expansion (\ref{expand-curv}), on the other hand, ensures that the affine curvature  ${\mathbb{R}}_{\mu\nu}(\Gamma)$ is equal to the metrical curvature $R_{\mu\nu}(g)$ up to a doubly-Planck suppressed remainder. In essence, what happened is that the affine dynamics took the affine curvature ${\mathbb{R}}$ from its UV value $\Lambda_\wp^2$  in (\ref{map}) to its IR value $R$ in (\ref{expand-curv}). 

Under the expansion of the affine curvature in \eqref{expand-curv}, 
the metric-Palatini action (\ref{action-1}) reduces to a metrical gravity theory
\begin{eqnarray}
     \int d^4x \sqrt{-g} \Bigg\{\! && - V_{\rm O}
 -\frac{{\mathbb{R}}(g)}{16\pi G} - \frac{c_{\rm O}}{16} \left({\mathbb{R}}(g)\right)^2 +\frac{c_S}{4}  S^\dagger S\, {\mathbb{R}}(g) + c_V V^{\mu}\!\left({\mathbb{R}}_{\mu\nu}(\Gamma)- R_{\mu\nu}({}^g\Gamma)\right)\!V^{\nu}\! \Bigg\}\nonumber\\
&&\xrightarrow{\rm equation\, (\ref{expand-curv})}
\int d^4x \sqrt{-g} \Bigg\{\!
- V_{\rm O}-\frac{R}{16\pi G}   - \frac{c_{\rm O}}{16} R^2  -\frac{c_S}{4} S^\dagger S  R + {\mathcal{O}}\!\left(G\right)\!\Bigg\}
\label{reduce-nongauge}
\end{eqnarray}
in which the loop factors $c_S$ and $c_V$ depend on the QFT (calculated for the SM in \cite{demir1,demir2}).  The loop factor $c_{\rm O}$, associated with the quartic ($\Lambda^4$) corrections in the flat spacetime effective QFT in (\ref{action-0}), turned to the coefficient of quadratic-curvature $(R^2)$ term in the symmergent GR action in (\ref{reduce-nongauge}). At one loop, it takes the value
\begin{eqnarray}
\label{param1}
 c_{\rm O} = \frac{n_b-n_f}{128 \pi^2}
\end{eqnarray}
in which  $n_b$ ($n_f$) stands for the total number of bosonic (fermionic) degrees of freedom in the underlying QFT (including the color degrees of freedom). Both the $n_b$ bosons and $n_f$ fermions contain not only the known standard model particles  but also the completely new particles. 

The dynamical evolution up to \eqref{reduce-nongauge} gives rise to a new framework in which {\it (i)} gauge symmetries get restored (no $V_\mu (\dots)V_\nu$ terms in \eqref{reduce-nongauge} anymore), 
{\it (ii)} the gravity sector is given by the Einstein-Hilbert term plus a curvature-squared term, and {\it (ii)}  matter sector is a renormalized QFT with new particles beyond the SM (which do not have to interact non-gravitationally with the SM particles). We call this new framework gauge symmetry-restoring emergent gravity or simply {\it symmergent gravity} to distinguish is from other emergent or induced gravity theories in the literature.  

The vacuum energy density $V_{\rm O}$ in the symmergent action (\ref{reduce-nongauge}) belongs to the non-power-law sector of the flat spacetime effective QFT. At one loop,  it takes the value 
\begin{eqnarray}
\label{param2}
 V_{\rm O} = \frac{{\rm str}\left[{\mathcal{M}}^4\right]}{64 \pi^2}
\end{eqnarray}
after discarding a possible tree-level contribution. Being a loop-induced quantity,  Newton's constant in (\ref{MPl}) involves super-trace of  $({\rm masses})^2$ of the QFT fields. In this regard, the potential energy $V_{\rm O}$, involving the super-trace of  $({\rm masses})^4$ of the QFT fields, is expected to expressible in terms of $G$. To this end, it proves useful to the technique used in \cite{reggiealiben} in which one starts with mass degeneracy limit in which each and every boson and fermion possess equal masses, $m_b=m_f=M_0$, for all $b$ and $f$. (Here, $M_0$ is the mean value of all the field masses.) Under this degenerate mass spectrum the potential $V_{\rm O}$ simplifies as follows
\begin{eqnarray}
\label{analyze-VO-1}
V_{\rm O} = \frac{{\rm str}\left[{\mathcal{M}}^4\right]}{64 \pi^2} = \frac{1}{64\pi^2}\left(\sum_{\rm B} m^4_{\rm B}- \sum_{\rm F} m^4_{\rm F}\right)\xrightarrow{\rm mass\,  degeneracy}\frac{M_0^4}{64\pi^2}(n_b - n_f)=\frac{M_0^2}{8\pi G}=\frac{1}{2(8\pi G)^2 c_{\rm O}}
%\nonumber\\ &\approx&  \frac{1}{64\pi^2} \left(\left(\sum_{\rm B} m^2_{\rm B}\right)^2 - \left(\sum_{\rm F} m^2_{\rm F}\right)^2\right)\nonumber\\ &=& \frac{1}{64\pi^2} \left(\sum_{\rm B} m^2_{\rm B} - \sum_{\rm F} m^2_{\rm F}\right) \left(\sum_{\rm B} m^2_{\rm B} + \sum_{\rm F} m^2_{\rm F}\right)\\ &=& \frac{1}{8\pi G} \left(\sum_{\rm B} m^2_{\rm B} + \sum_{\rm F} m^2_{\rm F}\right) \nonumber
\end{eqnarray}
after using the definitions of $G$ in (\ref{MPl}) and $c_{\rm O}$ formula in (\ref{param1}). Now, a non-degenrate QFT mass spectrum may be represented by reparametrizing the potential energy \eqref{analyze-VO-1} as
\begin{eqnarray}
\label{analyze-VO-2}
V_{\rm O} &=& \frac{1-\alpha}{(8\pi G)^2 c_{\rm O}} 
\end{eqnarray}
in which the new parameter $\alpha$ is introduced as a measure of the deviations of the boson and fermion masses from the QFT characteristic scale $M_0$. Clearly, $\alpha=1/2$ corresponds to the degenerate case in (\ref{analyze-VO-1}). Alternatively,  $\alpha=1$ represents the case in which $\sum_{\rm B} m^4_{\rm B}= \sum_{\rm F} m^4_{\rm F}$ in (\ref{analyze-VO-1}). In general,  $\alpha>1$ ($\alpha<1$) corresponds to the boson (fermion) dominance in terms of the trace $({\rm masses})^4$.

A  glance at the second line of (\ref{reduce-nongauge}) reveals that symmergent gravity is an $R+R^2$ gravity theory with non-zero cosmological constant. In fact, it can be put in the form  
\begin{eqnarray}
S=\frac{1}{16 \pi G} \int d^4 x \sqrt{-g}\left(R+f(R)\right)
\label{fr-action}
\end{eqnarray}
after discarding the scalars $S$ and the other matter fields, switching to $(-,+,+,+)$ metric signature, and introducing the curvature function
\begin{eqnarray}
f(R) =- \pi G c_{\rm O} R^2 -16 \pi G V_{\rm O}
\label{f(R)}
\end{eqnarray}
as a combination of the quadratic curvature term and the vacuum energy.
The Einstein field equations arising from 
the action (\ref{fr-action}) 
\begin{eqnarray} 
\label{f1}
\mathit{R}_{\mu \nu} (1+f^\prime(R))-\frac{1}{2} g_{\mu \nu} (R + f(R)) + (\nabla_\mu \nabla_\nu - \Box g_{\mu\nu})f^\prime(R) = 0
\end{eqnarray}
take the form $R_0(f^\prime(R_0)-1) - 2f(R_0) = 0$  upon contraction at constant curvature ($R=R_0$). This equation leads to the solution
\begin{eqnarray}
R_0 = 32\pi G V_{\rm O} = \frac{1-\alpha}{2\pi G c_{\rm O}}
\label{R0-gecici}
\end{eqnarray}
after using the vacuum energy formula in (\ref{analyze-VO-2}) in the second equality. It is clear that this constant-curvature solution exists only if the vacuum energy is nonzero.  Indeed, for a quadratic gravity with $f(R)= b R^2$ one gets the solution $R_0=0$. But for a quadratic gravity like $f(R)=\hat a+bR^2$ one finds $R_0=-2 \hat a \neq 0$, which reduces to the $R_0$ in (\ref{R0-gecici}) for $\hat a=-16 \pi G V_{\rm O}$. 

The Einstein field equations (\ref{f1}) possess 
static, spherically-symmetric, constant-curvature solutions of the form  
\cite{javlonla,irfanaliben,reggiealiben}
\begin{eqnarray}
\label{metric}
ds^2=-f(r) dt^2 + \frac{dr^2}{f(r)} + r^2 (d\theta^2+ \sin^2\theta d\phi^2)
\end{eqnarray}
in which
\begin{eqnarray} 
\label{smetric}
    f(r)=1-\frac{2 G M}{r}-\frac{(1-\alpha)}{24\pi G c_{\rm O}} r^2
\end{eqnarray}
is the lapse function following from (\ref{R0-gecici}). {\color{black} Looking at the $f(r)$ expression, one concludes that it is essentially the standard lapse function with cosmological constant. This actually is not surprising since $R=$``constant" solution  is similar in structure to $f(R)=$``vacuum energy" solution. In spite of this formal similarity, however, the symmergent lapse function $f(r)$ possesses the important feature that it is directly sensitive to the matter sector (masses of scalars, fermions, vectors in the underlying QFT) via the quadratic curvature coefficient $c_O$ in symmergent gravity. Besides, under the simplifying analyses that led to \eqref{analyze-VO-2}, the symmergent $f(r)$ is sensitive also to the vacuum energy (super-trace over quadratics of the particle masses) via the parameter $\alpha$. Hence, in the sequel, we will analyze this symmergent 
constant-curvature configuration ($R=R_0\neq 0$) in both the dS ($V_{\rm O}>0$ namely $\alpha<1$) and AdS ($V_{\rm O}<0$ namely $\alpha>1$) spacetimes. Our analyses of the QNMs and greybody factors will put constraints on the symmergent parameters.}

\section{Perturbations and Quasinormal modes}\label{sec3} 
In this section, we shall study massless scalar and massless vector perturbations about the Schwarzschild-dS/AdS metric in \eqref{metric}. To this end, we assume that the test field \textit{i.e.} (a scalar field or a vector field) or has negligible impact on
the black hole spacetime (negligible back reaction) \cite{lopez2020,Chandrasekhar_qnms}. To obtain the QNMs, we derive Schr\"odinger-like wave equations for each case by considering the
corresponding conservation relations on the concerned spacetime. The
equations will be of the Klein-Gordon type for scalar fields and the Maxwell
equations for electromagnetic fields. To calculate the QNMs, we
use two different methods \textit{viz.}, \textit{Asymptotic Iteration Method} and
\textit{Pad\'e averaged 6th order WKB approximation method}.

Taking into account the axial perturbation only, one may write the perturbed
metric in the following way \cite{lopez2020} 
\begin{equation}  \label{pert_metric}
ds^2 = -g_{tt} dt^2 + r^2 \sin^2\!\theta\, (d\phi - p_1(t,r,\theta)\,
dt - p_2(t,r,\theta)\, dr - p_3(t,r,\theta)\, d\theta)^2 + g_{rr}\, dr^2 +
r^2 d\theta^2,
\end{equation}
here the parameters $p_1, p_2$ and $p_3$ define the perturbation introduced
to the black hole spacetime. The metric functions $g_{tt}=f(r)$ and $g_{rr}=1/f(r)$ serve as the zeroth order (static and spherically-symmetric) terms  in the perturbative expansion in $p_i$.

\subsection{Massless Scalar Perturbation}
\label{massless-scalar}
We start by taking into account a massless scalar field in the vicinity of
the symmergent black hole in \eqref{metric}. Since the massless scalar $\Phi$ is assumed to have a negligible back reaction, its motion equation $\square \Phi = 0$ takes the form 
\begin{equation}  \label{scalar_KG}
\dfrac{1}{r^2 \sqrt{g_{tt} g_{rr}^{-1}}} \left[r^2 \sqrt{g_{tt} g_{rr}^{-1}}\,
\Phi_{,r} \right]_{,r} +  \dfrac{g_{tt}}{r^2\sin\theta}\left(\sin\theta
\Phi_{,\theta}\right)_{,\theta}+ \dfrac{g_{tt}}{r^2 \sin^2\theta} (\Phi)_{,\phi\phi}- (\Phi)_{,tt} =0
%\sqrt{g_{rr}}\, \mathcal{F}_{,tt} = 0,
%
%\square \Phi = \dfrac{1}{\sqrt{-g}} \partial_\mu (\sqrt{-g} g^{\mu\nu}
%\partial_\nu \Phi) = 0
\end{equation}
in the symmergent black hole spacetime in \eqref{metric}. In this Klein-Gordon equation, the structure of the angular part suggests that polar orientation basis functions are the associated Legendre polynomials $P_l^m(\cos\theta)$ since they obey 
\cite{sommerfeld,chandrasekhar}
\begin{eqnarray}
\dfrac{1}{\sin\theta} \dfrac{d}{d\theta}\left(\sin\theta \frac{d P_l^m(\cos\theta)}{d\theta}\right)-\frac{m^2}{\sin^2\theta}\Phi = - l (l+1)  P_l^m(\cos\theta)
\label{compare-with}
\end{eqnarray}
after separating $\phi$ part by letting $\partial^2_\phi \Phi = -m^2 \Phi$. Then, in this basis, at a given radius $r$, the scalar perturbation can be expanded in multipoles  in the form
\begin{equation}
\Phi(t,r,\theta, \phi) = \dfrac{1}{r} \sum_{l,m} \sqrt{\dfrac{(2 l +1)}{4\pi} \dfrac{(l-m)!}{(l+m)!}} \psi_{sl}(t,r) e^{im \phi}
P_l^m (\cos\theta)
\label{Phi-expand}
\end{equation}
over all values of the polar index $l$ and azimuth index $m$ \cite{sommerfeld,chandrasekhar}. The function $\psi_{sl}(t,r)$ is the radial time-dependent wave
function. The use of the expansion (\ref{Phi-expand}) in the Klein-Gordon equation in (\ref{scalar_KG}) leads to the stationary Schr\"odinger equation
\begin{equation}  \label{radial_scalar}
\partial^2_{r_*} \psi(r_*)_{sl} + \omega^2 \psi(r_*)_{sl} = V_s(r) \psi(r_*)_{sl},
\end{equation}
after defining the tortoise coordinate $r_*$ as 
\begin{equation}  \label{tortoise}
\dfrac{dr_*}{dr} = \sqrt{g_{rr} g_{tt}^{-1}} = f(r)
\end{equation}
such that $V_s(r)$ stands for the effective potential  \cite{lopez2020}
\begin{equation}  \label{Vs}
V_s(r) = |g_{tt}| \left( \dfrac{l(l+1)}{r^2} +\dfrac{1}{r \sqrt{|g_{tt}|
g_{rr}}} \dfrac{d}{dr}\sqrt{|g_{tt}| g_{rr}^{-1}} \right) = f(r) \left( \dfrac{l(l+1)}{r^2} +\dfrac{1}{r} \dfrac{d}{dr}f(r) \right)
\end{equation}
so that, now, $l$ becomes the multipole moment of the black hole's
QNMs. The frequency $\omega$ in the stationary Schr\"odinger equation \eqref{radial_scalar} is defined via $\partial_t^2 \psi_{sl} (t,r) =-\omega^2 \psi_{sl}(t,r)$ and this definition is made possible by the diagonal and static nature of the symmergent metric in \eqref{metric}. The wavefunction $\psi(r_*)_s$ stands for scalar QNMs \cite{Chandrasekhar_qnms,lopez2020}. 

\subsection{Massless Vector Perturbation}
\label{massless-vector}
Having done with the scalar perturbations, now we move to the massless vector perturbations, which are nothing but the electromagnetic perturbations. In analyzing vector perturbations it proves useful to use the tetrad formalism  \cite{Chandrasekhar_qnms,lopez2020} in which the curved metric $g_{\mu\nu}$ in \eqref{metric} is projected on the flat metric $\eta_{\bar \mu \bar \nu}$ via the vierbein $e_\mu^{\bar \mu}$ in the form $g_{\mu\nu}= e_\mu^{\bar \mu}\eta_{\bar \mu \bar \nu}e^{\bar \nu}_\nu$. (Here, notation is such that while $\mu, \nu, \dots$ refer to curved spacetime, $\bar \mu,  \bar \nu, \dots$ refer to flat spacetime.) The vierbeins satisfy the relations $e^{\bar \mu}_\mu e^\mu_{\bar \nu} = \delta^{\bar \mu}_{\bar \nu}$, $e^{\bar \mu}_\mu e^\nu_{\bar \mu} = \delta^{\nu}_{\mu}$ and $e^{\bar \mu}_\mu = g_{\mu\nu} \eta^{\bar\mu \bar\nu} e^\nu_{\bar\nu}$ so that a given vector $S_\mu$ and a tensor $P_{\mu\nu}$ can be expressed as $S_\mu = e^{\bar \mu}_\mu S_{\bar\mu}$ (with the inverse $S_{\bar\mu} = e^\mu_{\bar\mu} S_\mu$) and $P_{\mu\nu} = e^{\bar\mu}_\mu e^{\bar\nu}_\nu P_{\bar\mu \bar\nu}$ (with the inverse
$P_{\bar\mu \bar\nu} = e^\mu_{\bar\mu} e^\nu_{\bar\nu} P_{\mu\nu}$). 

The Jacobi identity $\partial_{[\bar \mu} F_{\bar\nu \bar\rho]} = 0$ for the electromagnetic field strength $F_{\bar\mu\bar\nu}=-F_{\bar\nu \bar\mu}$ leads to two constraints
\begin{align}
\left( r \sqrt{g_{tt}}\, F_{\bar t\bar \phi}\right)_{,r} + r \sqrt{g_{rr}}\,
F_{\bar\phi \bar r, t} &=0,  \label{em1} \\
\left( r \sqrt{g_{tt}}\, F_{\bar t \bar\phi}\sin\theta\right)_{,\theta} + r^2
\sin\theta\, F_{\bar \phi \bar \theta, t} &=0.  \label{em2}
\end{align}
In addition to these two, there is a third relation
\begin{equation}  \label{em3}
\left( r \sqrt{g_{tt}}\, F_{\bar\phi \bar r}\right)_{,r} + \sqrt{g_{tt} g_{rr}}%
\, F_{\bar \phi \bar \theta,\theta} + r \sqrt{g_{rr}}\, F_{\bar t \bar \phi, t} = 0
\end{equation}
following from the electric charge conservation 
\begin{equation}
\eta^{\bar\mu \bar\nu}\partial_{\bar\mu} F_{\bar\nu \bar\lambda} =0
\end{equation}
in the $\bar \lambda = \bar \phi$ direction. Now, using the equations \eqref{em1} and \eqref{em2} in the time derivative of 
\eqref{em3} one gets a two-derivative motion equation
\begin{equation}  \label{em4}
\left[ \sqrt{g_{tt} g_{rr}^{-1}} \left( r \sqrt{g_{tt}}\, \mathcal{F}
\right)_{,r} \right]_{,r} + \dfrac{g_{tt} \sqrt{g_{rr}}}{r} \left( \dfrac{%
\mathcal{F}_{,\theta}}{\sin\theta} \right)_{,\theta}\!\! \sin\theta - r 
\sqrt{g_{rr}}\, \mathcal{F}_{,tt} = 0,
\end{equation}
after introducing $\mathcal{F} = F_{\bar t \bar \phi} \sin\theta$. The expansion in \eqref{Phi-expand} for scalar perturbations $\Phi$ cannot be used for vector perturbations $F_{\bar\nu \bar\lambda}$ because, first of all, equation \eqref{em4} does not involve $\phi$ coordinate and, second of all, $\theta$ dependence of \eqref{em4} is not in the form required by the Legendre polynomials $P_l^m(\cos\theta)$. In fact, structure of the second term in \eqref{em4} suggests that the correct basis functions are the Gegenbauer polynomials $P_l(\cos\theta|-1)$  since they obey \cite{sommerfeld,chandrasekhar}
\begin{eqnarray}
\sin\theta \dfrac{d}{d\theta}\left(\dfrac{1}{\sin\theta} \frac{d P_l(\cos\theta|-1)}{d\theta}\right) = - l (l-1)  P_l(\cos\theta|-1)
\end{eqnarray}
as the defining differential equation \cite{sommerfeld}. But $l(l-1)$ is not the true eigenvalue for the polar part.  To get the correct value, we write the vector wavefunction $\mathcal{F}_l(r,\theta,t)$ in terms of $dP_l(\cos\theta|-1)/d\theta$  and perform the multipole expansion 
\begin{equation}
\mathcal{F}(t,r,\theta, \phi) = \sum_{l} \psi_{evl}(t,r) \dfrac{dP_l(\cos\theta|-1)}{\sin\theta d\theta}
\label{V-expand}
\end{equation}
and find that equation \eqref{em4} above  takes the form 
\begin{equation}  \label{em5}
\left[ \sqrt{g_{tt} g_{rr}^{-1}} \left( r \sqrt{g_{tt}}\, \psi_{v l}
\right)_{,r} \right]_{,r} + \omega^2 r \sqrt{g_{rr}}\, \psi_{v l} -
\dfrac{1}{r} g_{tt} \sqrt{g_{rr}}\, l(l+1)\, \psi_{v l} = 0
\end{equation}
to be compared with the equation \eqref{compare-with} of the scalar perturbations. In this equation, the frequency $\omega$  follows from the variable separation $\partial_t^2 \psi_{el} (t,r) =-\omega^2 \psi_{el}(t,r)$ allowed by the diagonal and static nature of the symmergent metric in \eqref{metric}. (We denote the vector perturbations by $\psi_{el}(t,r)$ in analogy with electromagnetism although we deal with most general vector perturbations.)  Now, using the tortoise coordinate $r_*$ in \eqref{tortoise}, the vector perturbations are found to obey the stationary Schr\"odinger equation
\begin{equation}
\partial^2_{r_*} \psi_{el}(r_*) + \omega^2 \psi_{el}(r_*) = V_{e}(r) \psi_{el}(r_*),
\end{equation}
with the effective potential 
\begin{equation}  \label{Ve}
V_{e}(r) = g_{tt}\, \dfrac{l(l+1)}{r^2}=f(r) \dfrac{l(l+1)}{r^2}\,.
\end{equation}
This effective potential $V_{e}(r)$ of vector perturbations differs from the effective potential $V_s(r)$ of scalar perturbations by the presence of the derivative of $g_{tt}$ in the latter.  

In practice, one would identify the massless vector here with the electromagnetic field. But, in general, symmergent gravity predicts existence of a plethora of new particles, which do not have to interact with the known particles non-gravitationally \cite{demir0,demir1,demir2}. And, some of these particles could be light or massless. All this means that the symmergent particle spectrum can contain massless scalars as in Sec. \ref{massless-scalar} or massless (dark) photons as above. In the presence of such (presumably dark) particles, one expects gravitational wave emissions to change considerably. 

\subsection{Behaviour of Potentials}

Here we shall briefly examine the characteristics of the perturbation potential pertaining to the aforementioned black hole. As the nature of the potential is closely intertwined with the QNMs, one can obtain a preliminary understanding of the QNMs by analyzing the behavior of the potential.

\begin{figure}[!h]
\centerline{
   \includegraphics[scale = 0.8]{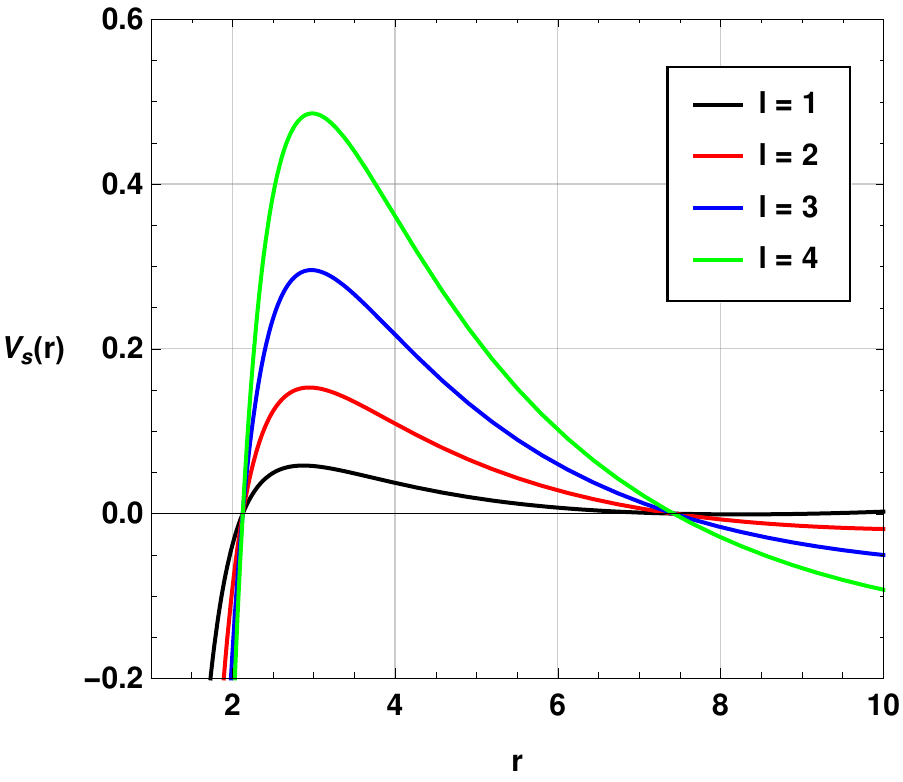}\hspace{0.5cm}
   \includegraphics[scale = 0.8]{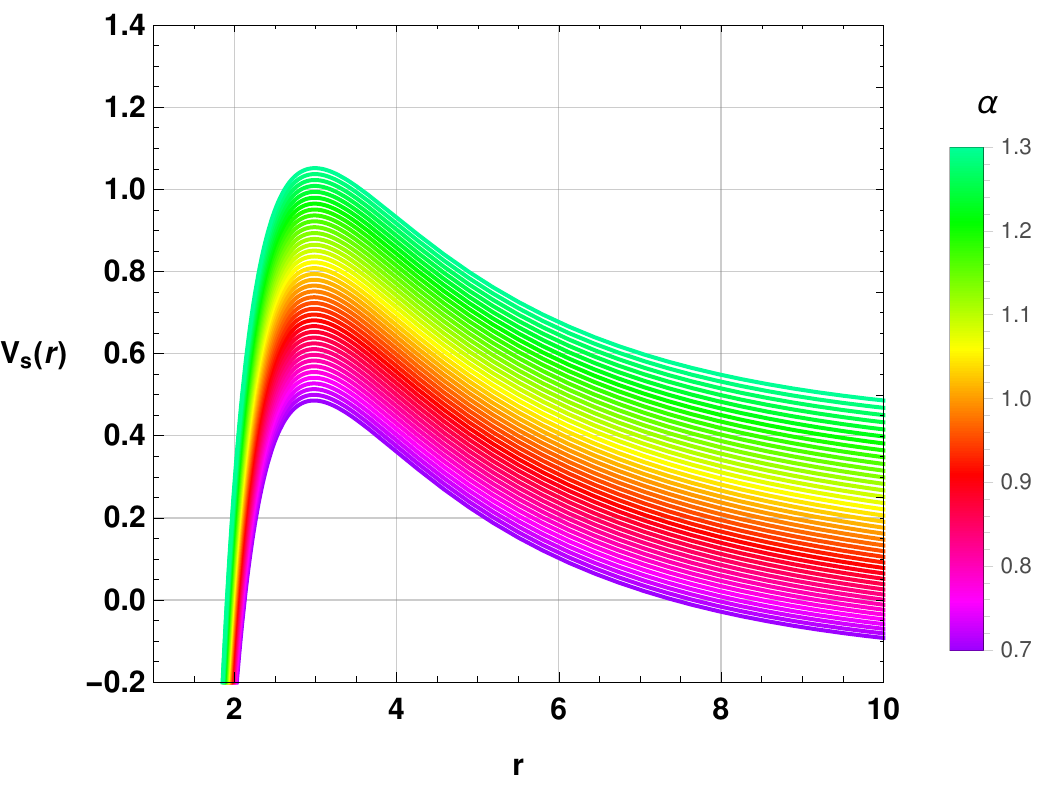}} \vspace{-0.2cm}
\caption{Variation of the scalar potential $V_s(r)$ with the radial distance $r$ for different values of the multipole moment $l$ with $M=1$, $G = 1$, $\alpha = 0.7$ and $c_{\rm O} = 0.3$ (left panel), and for different values of $\alpha$ with $l=4$, $M = 1$ and $c_{\rm O} = 0.3$ (right panel). }
\label{fig_Vs_01}
\end{figure}

\begin{figure}[!h]
\centerline{
   \includegraphics[scale = 0.8]{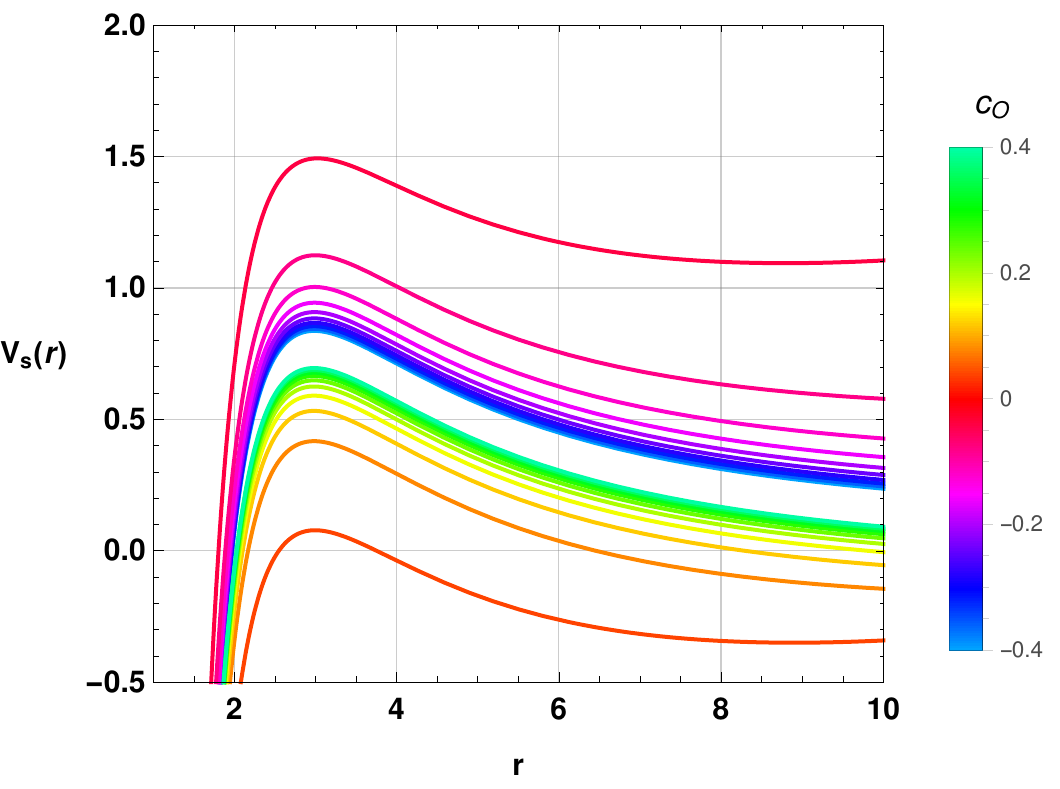}} \vspace{-0.2cm}
\caption{Variation of the scalar potential $V_s(r)$ with the radial distance $r$ for different values of $c_{\rm O}$ with $M=1$, $G = 1$, $l=4$ and $\alpha = 0.9$. }
\label{fig_Vs_02}
\end{figure}

\begin{figure}[!h]
\centerline{
   \includegraphics[scale = 0.8]{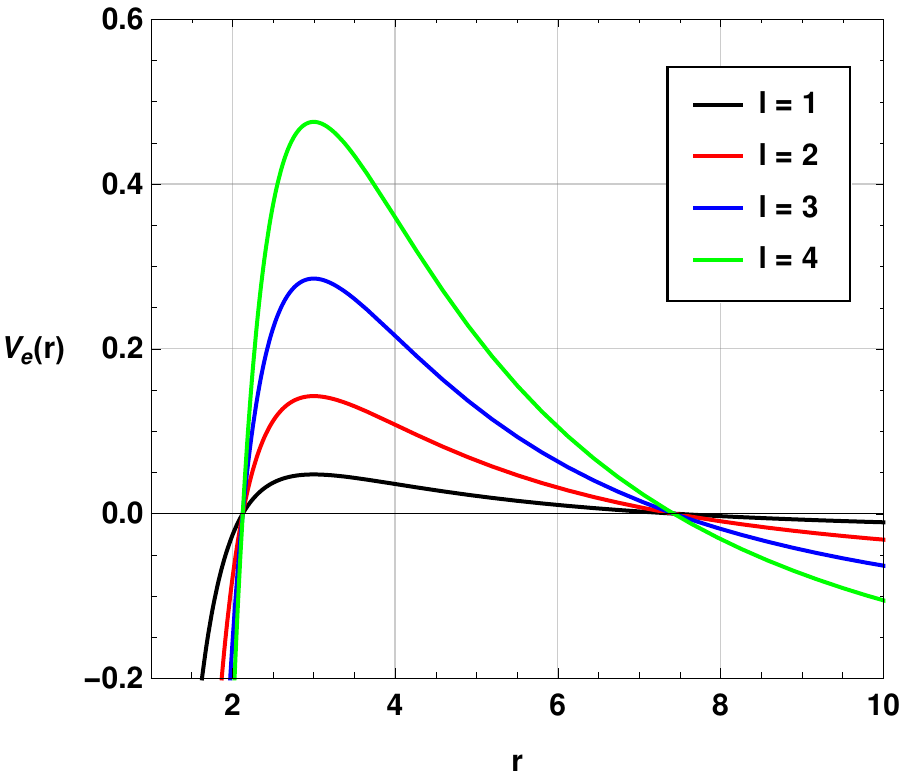}\hspace{0.5cm}
   \includegraphics[scale = 0.8]{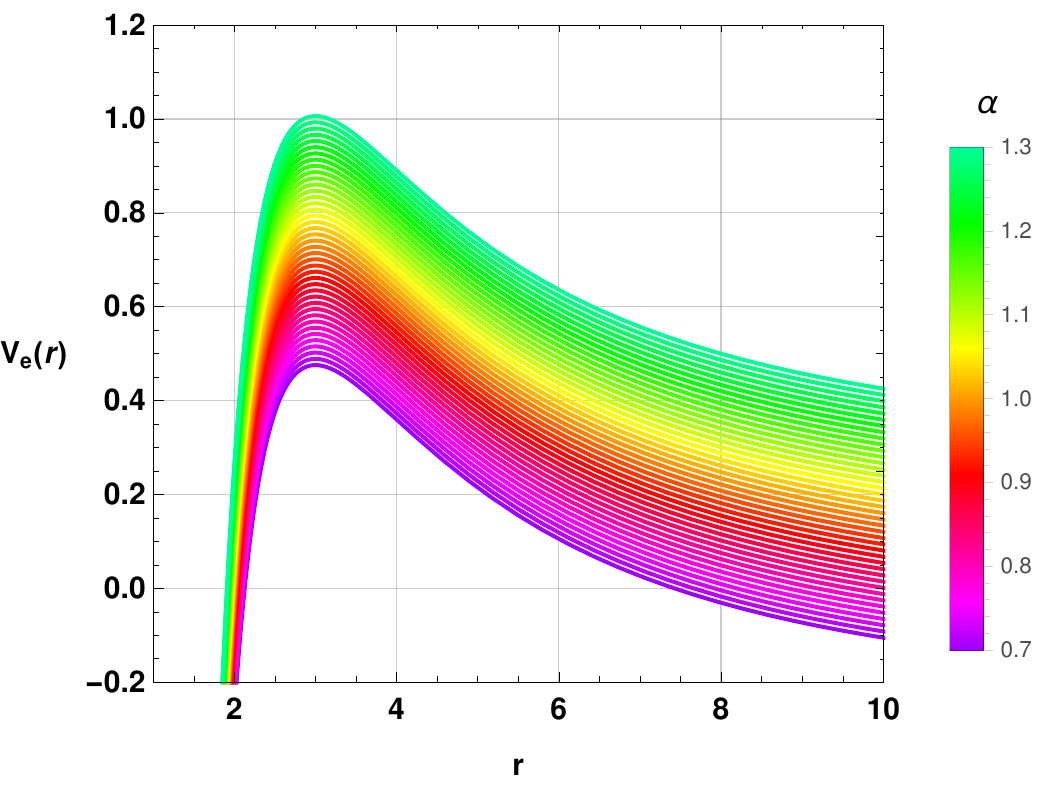}} \vspace{-0.2cm}
\caption{Variation of electromagnetic potential $V_e(r)$ with the radial distance $r$ for different values of the multipole moment $l$ with $M=1$, $G = 1$, $\alpha = 0.7$ and $c_{\rm O} = 0.3$ (left panel), and for different values of $\alpha$ with $M = 1$, $l=4$ and $c_{\rm O} = 0.3$ (right panel). }
\label{fig_Vem_01}
\end{figure}

\begin{figure}[!h]
\centerline{
   \includegraphics[scale = 0.8]{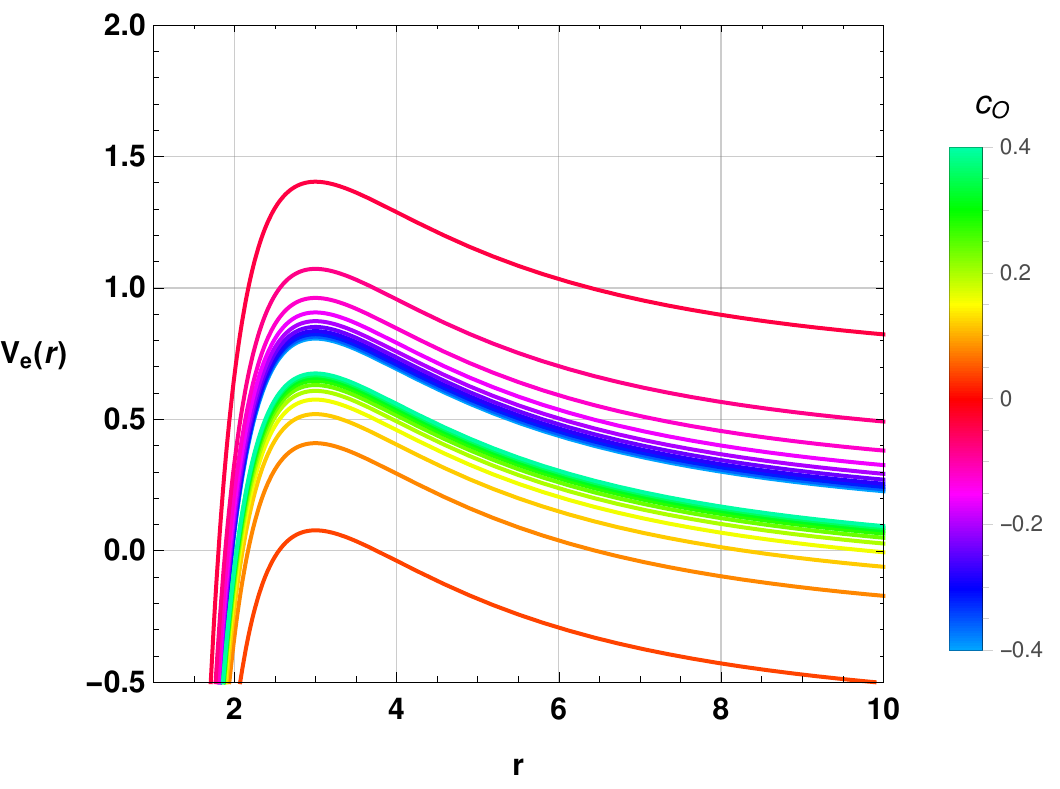}} \vspace{-0.2cm}
\caption{Variation of electromagnetic potential $V_e(r)$ with the radial distance $r$ for different values of $c_{\rm O}$ with $M = 1$, $l=4$ and $\alpha = 0.9$. }
\label{fig_Vem_02}
\end{figure}

On the first panel of Fig. \ref{fig_Vs_01}, we have plotted the scalar potential for different values of multipole moment $l$. The peak of the potential increases with an increase in the value of $l$ as expected. On the second panel of Fig. \ref{fig_Vs_01}, we have shown the variation of the scalar potential for different values of the model parameter $\alpha$. One can see that with an increase in the value of $\alpha$, the potential curve rises significantly. In Fig. \ref{fig_Vs_02}, we have shown the variation of the scalar potential with respect to the parameter $c_{\rm O}$. One may note that for very small positive values of $c_{\rm O}$, the potential curve increases rapidly and for very small negative values of $c_{\rm O}$, the potential curve decreases drastically. For larger negative values and larger positive values of $c_{\rm O}$, the potential curve moves towards a central value slowly.

We have plotted the similar curves for electromagnetic potential also in Figs. \ref{fig_Vem_01} and \ref{fig_Vem_02} respectively for different values of the model parameters and multipole moments. The qualitative behaviours of the electromagnetic potential curves are similar to those obtained for scalar potential. However, we have observed that the potential values for the scalar perturbation are higher than those corresponding to the electromagnetic one. As a result, we expect a similar behaviour in the QNMs spectrum also for both types of perturbations. However, these results suggest that the QNMs might have smaller values for the electromagnetic perturbation.

\subsection{The asymptotic iteration method}

The AIM (Asymptotic Iteration Method) is a powerful mathematical tool that is widely used for the numerical solution of differential equations, particularly for those that cannot be solved analytically. Its application is particularly significant in the study of QNMs in black holes and other systems with potential barriers \cite{AIM1, AIM2, AIM3, AIM4}. QNMs are characteristic oscillations that a system exhibits after a disturbance, and they play a crucial role in analyzing the stability and properties of black holes. The AIM employs a systematic iterative approach, which enables the derivation of accurate approximations to the QNMs by transforming the initial differential equation into a series of simpler equations that can be easily solved. This technique has achieved success in various physical systems and continues to be an active area of investigation.

%In the de Sitter or anti-de Sitter case, we define 
%$\mathcal{G} = 1/r$ following the Ref. \cite{AIM1}. 
Using definition $u = 1/r$ and following Ref.\ \cite{AIM1}, it is possible to write the master wave equation in the following form:
\begin{equation}
\frac{d^2 \psi}{d u^2} + \frac{\mathcal{Z}'}{\mathcal{Z}} \frac{d\psi} {d u} + \left[ \frac{\omega ^2-\mathcal{Z} \left(\left(1-s^2\right) \left(2 M G u-\frac{(\alpha -1) \left(s^2-4\right)}{48 \pi  c_{\rm O} G u^2}\right)+l (l+1) \right)}{\mathcal{Z}^2}\right]\!\psi = 0, \label{masterxi}
\end{equation}
where we have
\begin{equation}
\mathcal{Z}= \frac{\alpha -1}{24 \pi  c_{\rm O} G}-2 M G u^3+u^2.
\end{equation}
In the above expressions, $s=0$ represents scalar perturbation and $s=1$ represents electromagnetic perturbation. Now, to avoid the divergent characteristics of QNMs 
at the cosmological horizon, we define the wave function
$\psi(u)$ as
\begin{equation}
\psi(u) = e^{i\omega r_*}\, \mathcal{G} (u),  \label{SdScale}
\end{equation}
which allows us to rewrite Eq.\ \eqref{masterxi} as
\begin{equation}
\mathcal{Z} \mathcal{G}'' + (\mathcal{Z}'- 2 i\omega)\mathcal{G}' - \left[ \left(1-s^2\right) \left(2 M G u-\frac{(1-\alpha ) \left(4-s^2\right)}{48 \pi  c_{\rm O} G u^2}\right)+l (l+1) \right]\mathcal{G}=0. \label{youeq}
\end{equation}
From the appropriate quasinormal condition at the horizon of black hole $u_1$, we have
\begin{equation}
\mathcal{G}(r_*) =  (u-u_1)^{-\frac{i\omega}{\kappa_1}}\chi(r_*),
\end{equation}
where $\kappa_1$ is expressed as
\begin{equation}
\kappa_1 = \left.\frac 1 2 \frac{d f}{ dr}\right|_{r\to r_1}\!\!\!\! =M G u_1^2-\frac{1-\alpha }{24 \pi  c_{\rm O} G u_1}\,.
\end{equation}
In the aforementioned equations, the parameter $u_1$ is defined as the reciprocal of the event horizon radius of the black hole, denoted as $r_1$. By introducing the new function $\chi(r_*)$, Eq. \eqref{youeq} can be reformulated into a standard format suitable for the iterative process of the AIM when dealing with the differential equation. The reformulated equation takes the form:
\begin{eqnarray}
\chi''& = & \lambda_0(u)  \chi' + s_0(u) \chi\,,
\end{eqnarray}
where the parameters $\lambda_0$ and $s_0$ are defined as follows:
\begin{align}
\lambda_0(u) &= -\frac{1}{\mathcal{Z}} \left[\mathcal{Z}'- \frac{2i\omega }{ \kappa_1(u-u_1)} - 2 i\omega\right], \\[5pt]
s_0 (u) &=  \frac{1}{\mathcal{Z}} \left[\left(1-s^2\right) \left(2 M u-\frac{(1-\alpha ) \left(4-s^2\right)}{48 \pi  c_{\rm O} G u^2}\right)+l (l+1)\ +\frac{i \omega}{ \kappa_1(u-u_1)^2}\,\Big(\frac{i\omega}{\kappa_1} +1\Big) +(\mathcal{Z}'- 2 i\omega)\, \frac{i\omega}{ \kappa_1(u-u_1)}\right].
\end{align}
These new parameter definitions enable a more conventional representation of the equation, which facilitates the iterative process within the AIM framework.

To determine the scalar and electromagnetic QNMs for the symmergent black hole under concern, we shall employ the aforementioned differential equation. We shall adopt the the calculation methodology outlined in Ref. \cite{AIM1} in that we shall solve the differential equation \eqref{masterxi} using the techniques described in Ref. \cite{AIM1}. 

\subsection{WKB method with Pad\'e Approximation for Quasinormal modes}

In addition to the AIM, we shall utilize the WKB (Wentzel-Kramers-Brillouin) method, a widely recognized and established technique, to estimate the QNMs of the black holes. By comparing the results obtained via the two approaches, we aim to validate and strengthen our findings.

The WKB method was initially introduced by Schutz and Will \cite{Schutz} as a first-order approximation technique for determining QNMs. However, it is important to note that this method comes with inherent limitations, particularly a relatively higher degree of error. To overcome this drawback, researchers have subsequently developed higher-order versions of the WKB method, which have been widely employed in the analysis of black hole QNMs. Notable contributions in this area can be found in Refs. \cite{Will_wkb, Konoplya_wkb, Maty_wkb}.

Ref. \cite{Maty_wkb} suggested an improvement to the WKB technique by incorporating averaging of Pad\'e approximations. This advancement has proven to enhance the precision of QNM results, as reported in \cite{Konoplya_wkb}. Building upon these advancements, our investigation will employ the Pad\'e averaged 6th-order WKB approximation approach.

By incorporating the aforementioned refinements and utilizing the Pad\'e averaged 6th-order WKB approximation, we aim to obtain more accurate and reliable estimates for the QNMs of the black hole in this study. The comparative analysis with the previously employed method will serve as a critical step in validating the robustness of our findings and further advancing our understanding of black hole dynamics.

In Tables \ref{tab01} and \ref{tab02}, we have shown the QNMs obtained using AIM (up to 51 iterations) and Pad\'e averaged 6th order WKB approximation method. In 4th and 5th columns of the tables, we have shown the errors associated with the WKB approximation method, where the rms error is denoted by $\vartriangle_{rms}$ and another error term $\Delta_6$ is expressed as \cite{Konoplya_wkb} 
\begin{equation}
\Delta_6 = \dfrac{\vline \; \omega_7 - \omega_5 \; \vline}{2},
\end{equation}
here the terms $\omega_7$ and $\omega_5$ stand for the QNMs
calculated using $7$th and $5$th order Pad\'e averaged WKB approximation method.

\begin{table}[ht]
\caption{The QNMs of the symmergent black hole  for the massless scalar perturbation with $n= 0$, $M=1$, $G = 1$, $\alpha = 0.7$ and $c_{\rm O}=0.3$ using the AIM (up to 51 iterations) and using the  Pad\'e averaged 6th order WKB approximation method.}
\label{tab01}
\begin{center}
{\small 
\begin{tabular}{|cccccc|}
\hline
\;\;$l$ & \;\;AIM \;\; & \;\; Pad\'e averaged WKB\;\;
& $\vartriangle_{rms}$ & $\Delta_6$ & $\Delta_{AW}$ \\ \hline
$l=1$ & $0.2250819 - 0.0821453 i$ & $0.225068 - 0.0821502 i$ & $%
1.17414\times10^{-6}$ & $0.0000429369$ & $ 0.00608594\%$ \\ 
$l=2$ & $0.3813837 - 0.0788725 i$ & $0.381383 - 0.0788725 i$ & $%
4.91538\times10^{-8}$ & $1.72853\times10^{-6}$ & $ 0.000110976\%$ \\ 
$l=3$ & $0.5366843 - 0.0779927 i$ & $0.536684 - 0.0779928 i$ & $%
4.26502\times10^{-8}$ & $3.12754\times10^{-7}$ & $ 0.0000365382\%$ \\ 
$l=4$ & $0.6915178 - 0.0776355i$ & $0.691518 - 0.0776355 i$ & $%
1.4474\times10^{-8}$ & $8.59983\times10^{-8}$ & $ 0.0000250682\%$ \\ 
$l=5$ & $0.8461237 - 0.0774558i$ & $0.846124 - 0.0774557 i$ & $%
5.12218\times10^{-9}$ & $1.29086\times10^{-8}$ & $ 0.0000237887\%$ \\ \hline
\end{tabular}
}
\end{center}
\end{table}

\begin{table}[ht]
\caption{The QNMs of the symmergent black hole  for the electromagnetic (massless vector) perturbation with $n= 0$, $M=1$, $G = 1$, $\alpha = 0.7$ and $c_{\rm O}=0.3$ using the AIM (up to 51 iterations) and using the  Pad\'e averaged 6th order WKB approximation method.}
\label{tab02}
\begin{center}
{\small 
\begin{tabular}{|cccccc|}
\hline
\;\;$l$ & \;\;AIM \;\; & \;\; Pad\'e averaged WKB\;\;
& $\vartriangle_{rms}$ & $\Delta_6$ & $\Delta_{AW}$ \\ \hline
$l=1$ & $0.2008983 - 0.0748319i$ & $0.200881 - 0.0748496 i$ & $0.0000103311$
& $6.97624\times10^{-6}$ & $ 0.0114363\%$ \\ 
$l=2$ & $0.3677672 - 0.0763503i$ & $0.367767 - 0.0763516 i$ & $%
5.44868\times10^{-7}$ & $4.30915\times10^{-7}$ & $ 0.000387774\%$ \\ 
$l=3$ & $0.5271103 - 0.0767227i$ & $0.52711 - 0.0767229 i$ & $%
7.75823\times10^{-8}$ & $8.65772\times10^{-8}$ & $ 0.0000456533\%$ \\ 
$l=4$ & $0.6841178 - 0.0768714i$ & $0.684118 - 0.0768714 i$ & $%
1.70182\times10^{-8}$ & $3.86429\times10^{-8}$ & $ 0.000023411\%$ \\ 
$l=5$ & $0.8400880 - 0.0769457i$ & $0.840088 - 0.0769457 i$ & $%
5.24887\times10^{-9}$ & $1.37718\times10^{-8}$ & $ 0.0000248097\%$ \\
\hline
\end{tabular}
}
\end{center}
\end{table}

In Table \ref{tab01}, we have presented the QNMs of the massless scalar perturbation using the following model parameters: $c_{\rm O}= 0.3$, $\alpha = 0.7$, $G = M = 1$, and overtone number $n=0$. The term $\Delta_{AW}$ represents the discrepancy expressed as a percentage between the QNMs obtained through the AIM and the 6th order Pad\'e averaged WKB approximation method. It is evident that for $l=1$, $\Delta_{AW}$ exhibits a relatively higher value. Moreover, the Pad\'e averaged WKB method is also associated with larger rms errors. As the multipole moment $l$ increases, both the percentage deviation and the errors decrease significantly. These variations in errors can be attributed to the behavior of the WKB method, which produces less accurate outcomes when the difference between $l$ and $n$ is smaller. It is evident from both methods that the quasinormal frequencies and the damping rate augment with the increasing value of the multipole moment $l$.

Table \ref{tab02} presents the QNMs for electromagnetic perturbation at various values of $l$, using $c_{\rm O}= 0.3$, $\alpha = 0.7$, $G = M = 1$, and overtone number $n=0$. Similarly to the previous case, the deviation term $\Delta_{AW}$ shows higher values for smaller multipole moments $l$. However, as the value of $l$ increases, both the deviation and the error terms decrease significantly.

Comparing both tables, it is evident that the quasinormal frequencies and damping rate are lower for electromagnetic perturbation than for massless scalar perturbation.

\begin{figure}[htbp]
\centerline{
   \includegraphics[scale = 0.5]{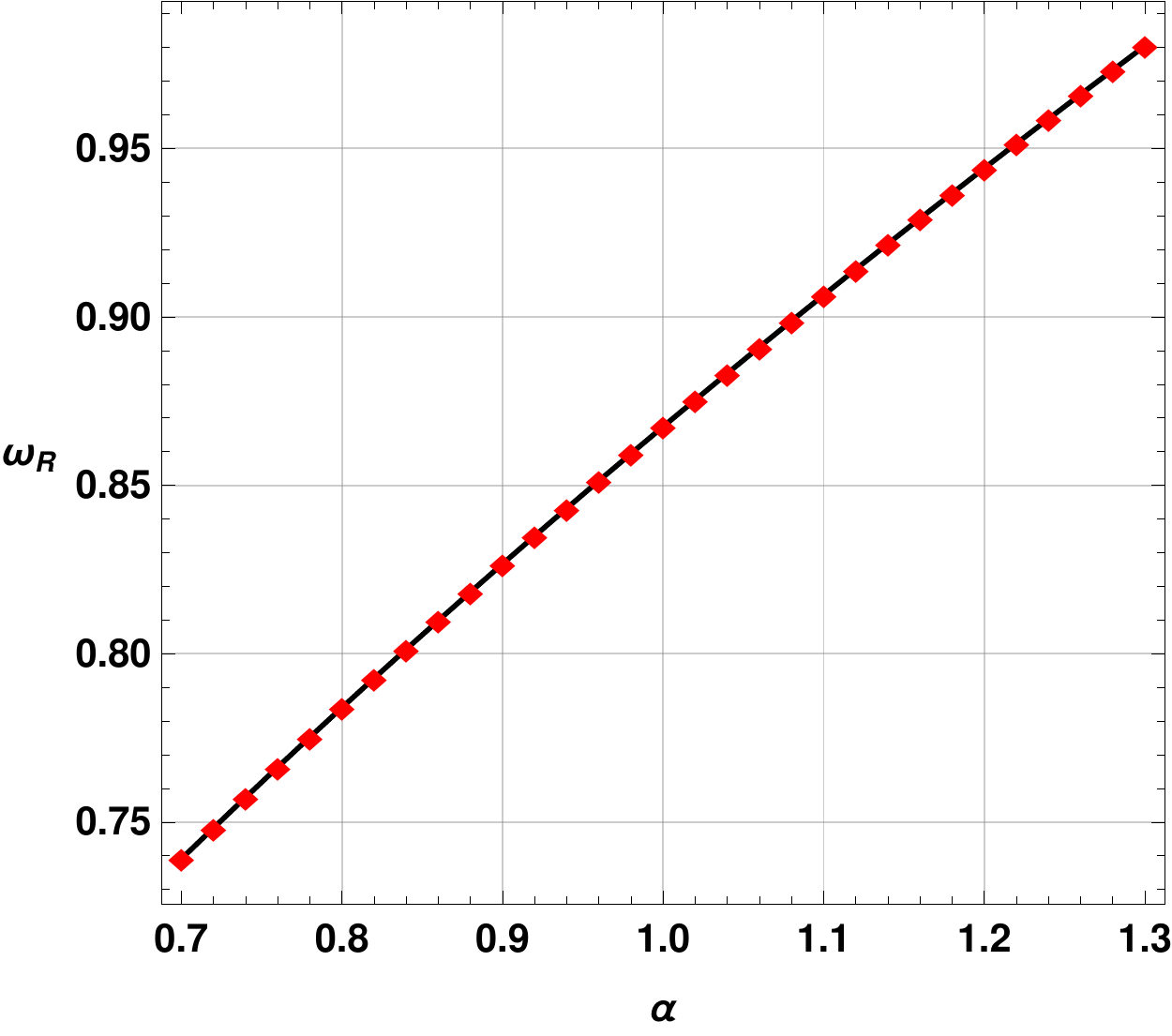}\hspace{0.5cm}
   \includegraphics[scale = 0.5]{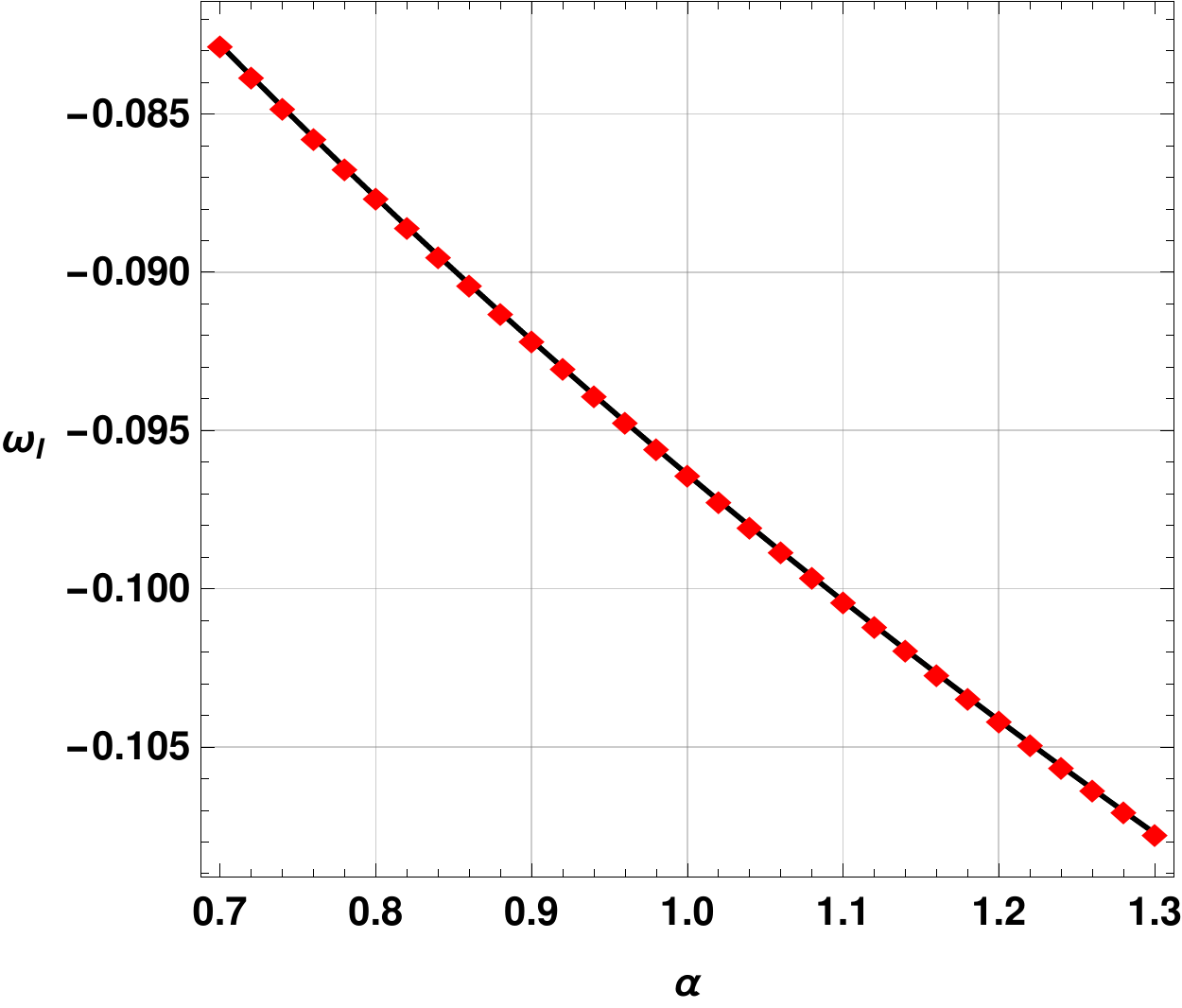}} \vspace{-0.2cm}
\caption{The real (left panel) and imaginary (right panel) parts of the QNMs of the symmergent black hole for massless scalar perturbations as a function of the vacuum energy parameter $\alpha$ with $M=1$, $G = 1$, $n= 0$, $l=4$ and $c_{\rm O} = 0.4$.}
\label{QNMs01}
\end{figure}

\begin{figure}[htbp]
\centerline{
   \includegraphics[scale = 0.5]{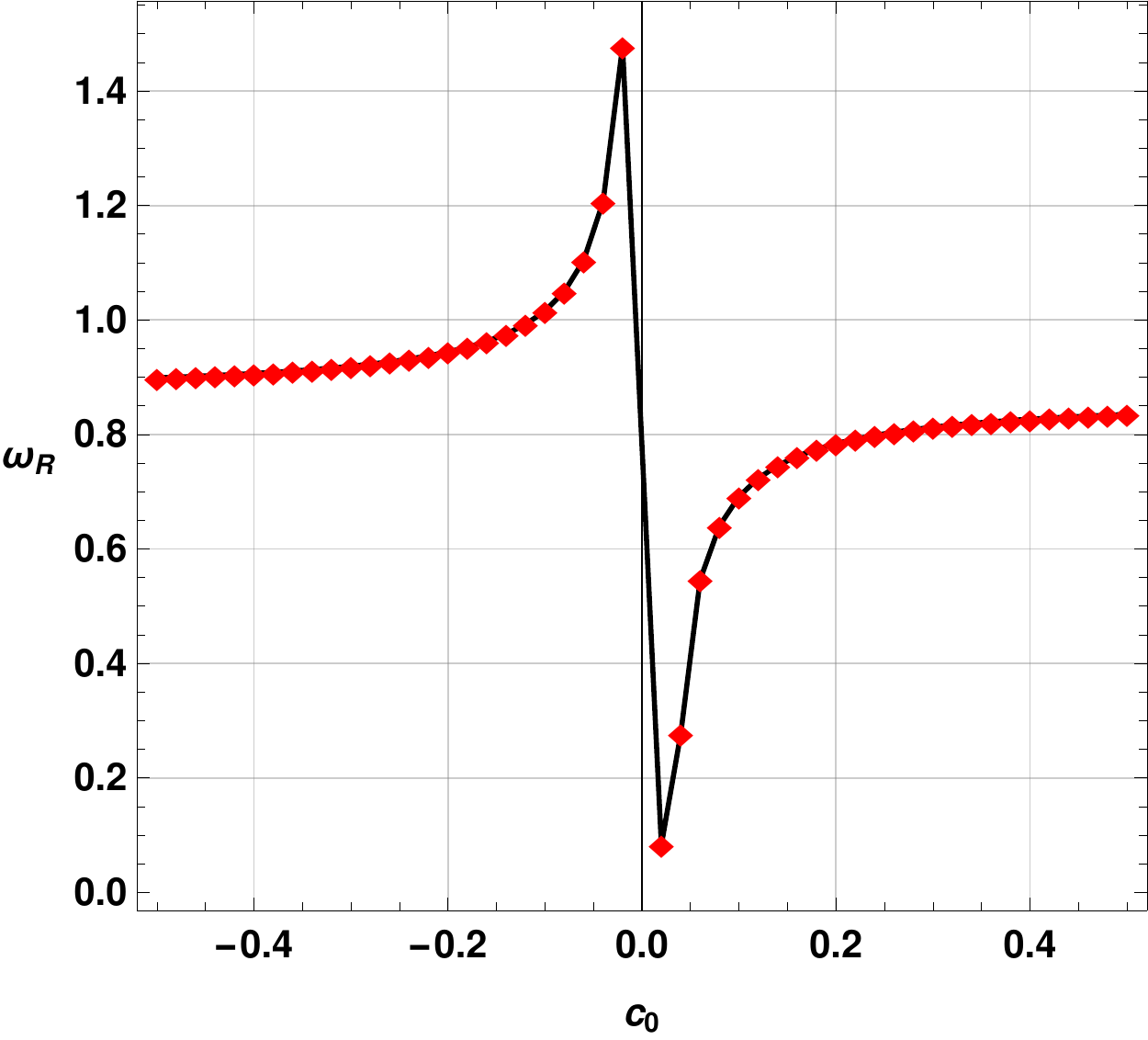}\hspace{0.5cm}
   \includegraphics[scale = 0.5]{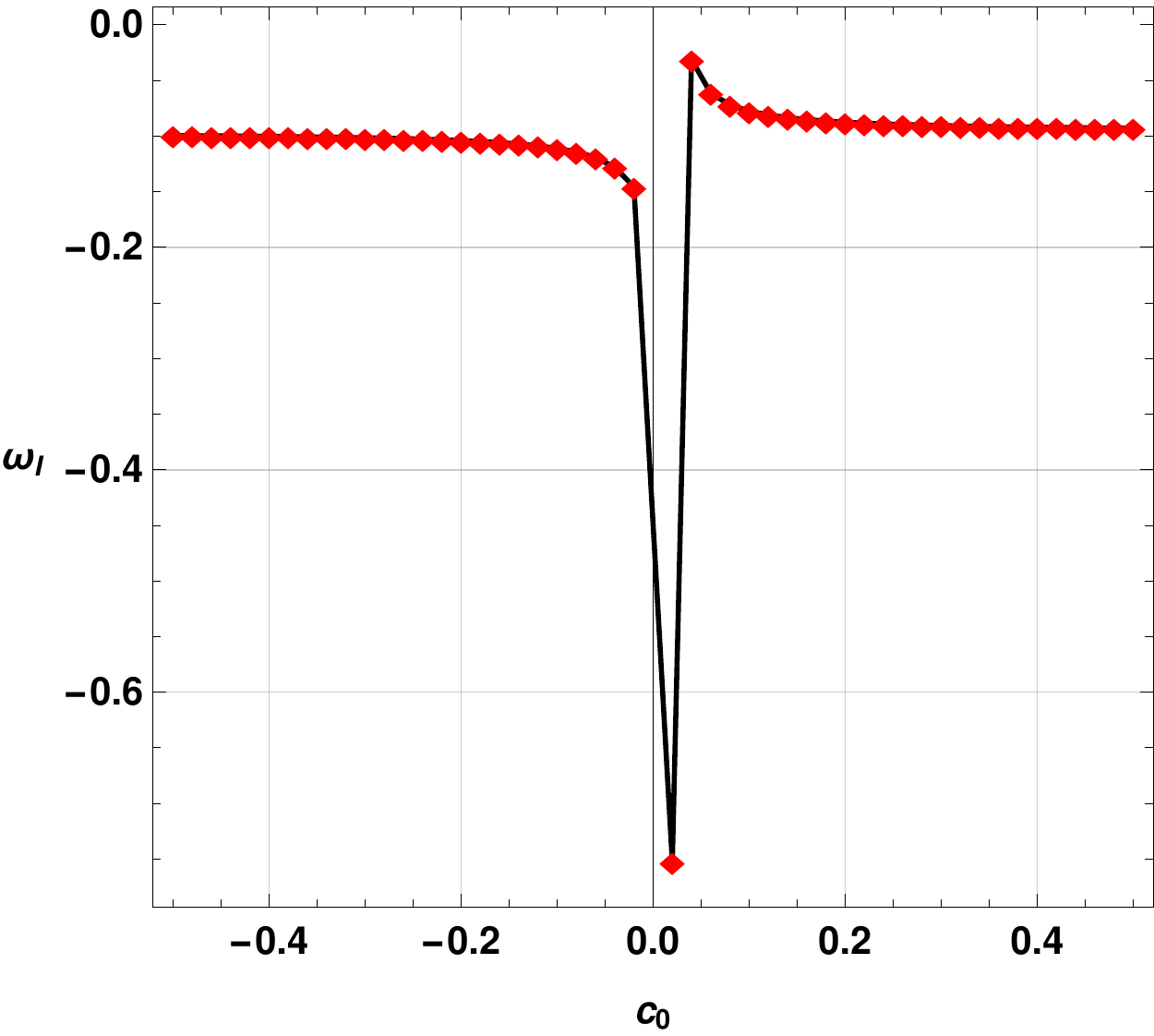}} \vspace{-0.2cm}
\caption{The real (left panel) and imaginary (right panel) parts of the QNMs of the symmergent black hole for massless scalar perturbations as a function of the symmergent parameter $c_{\rm O}$ with $M=1$, $G = 1$, $n= 0$, $l=4$ and $\alpha = 0.9$.}
\label{QNMs02}
\end{figure}

\begin{figure}[htbp]
\centerline{
   \includegraphics[scale = 0.5]{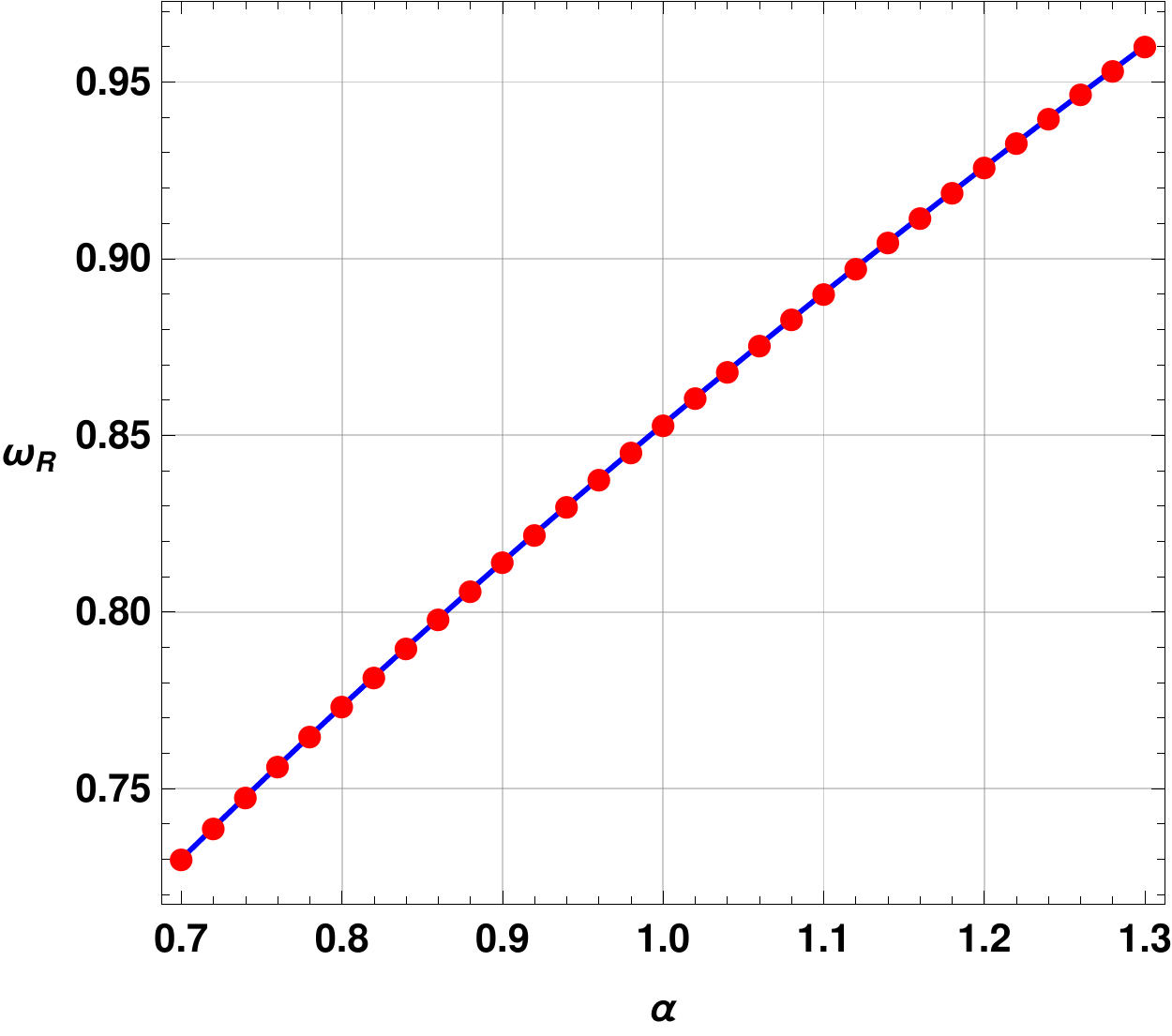}\hspace{0.5cm}
   \includegraphics[scale = 0.5]{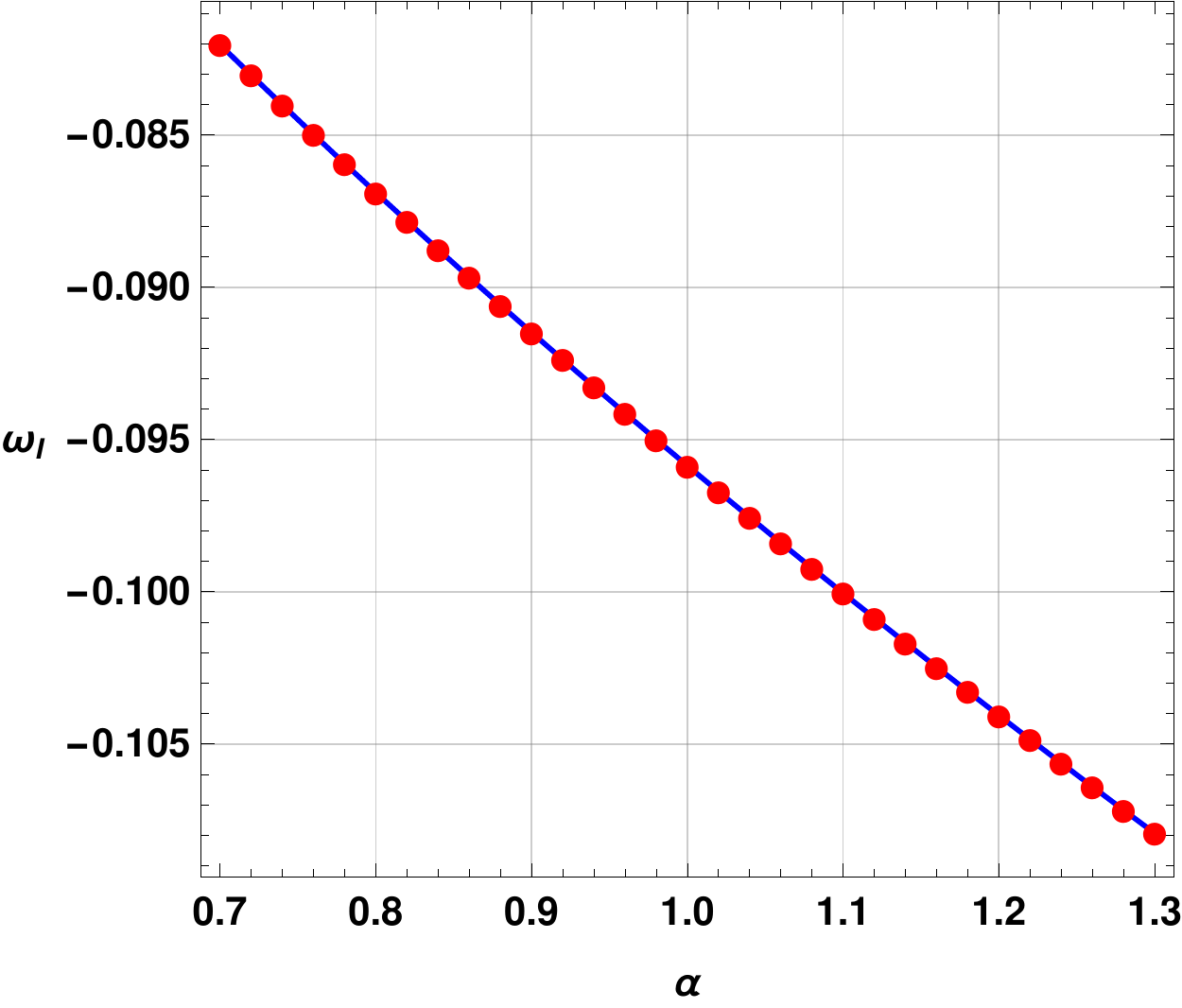}} \vspace{-0.2cm}
\caption{The real (left panel) and imaginary (right panel) parts of the QNMs of the symmergent black hole for electromagnetic perturbations as a function of the vacuum energy parameter $\alpha$ with $M=1$, $G = 1$, $n= 0$, $l=4$ and $c_{\rm O} = 0.4$.}
\label{QNMs03}
\end{figure}

\begin{figure}[htbp]
\centerline{
   \includegraphics[scale = 0.5]{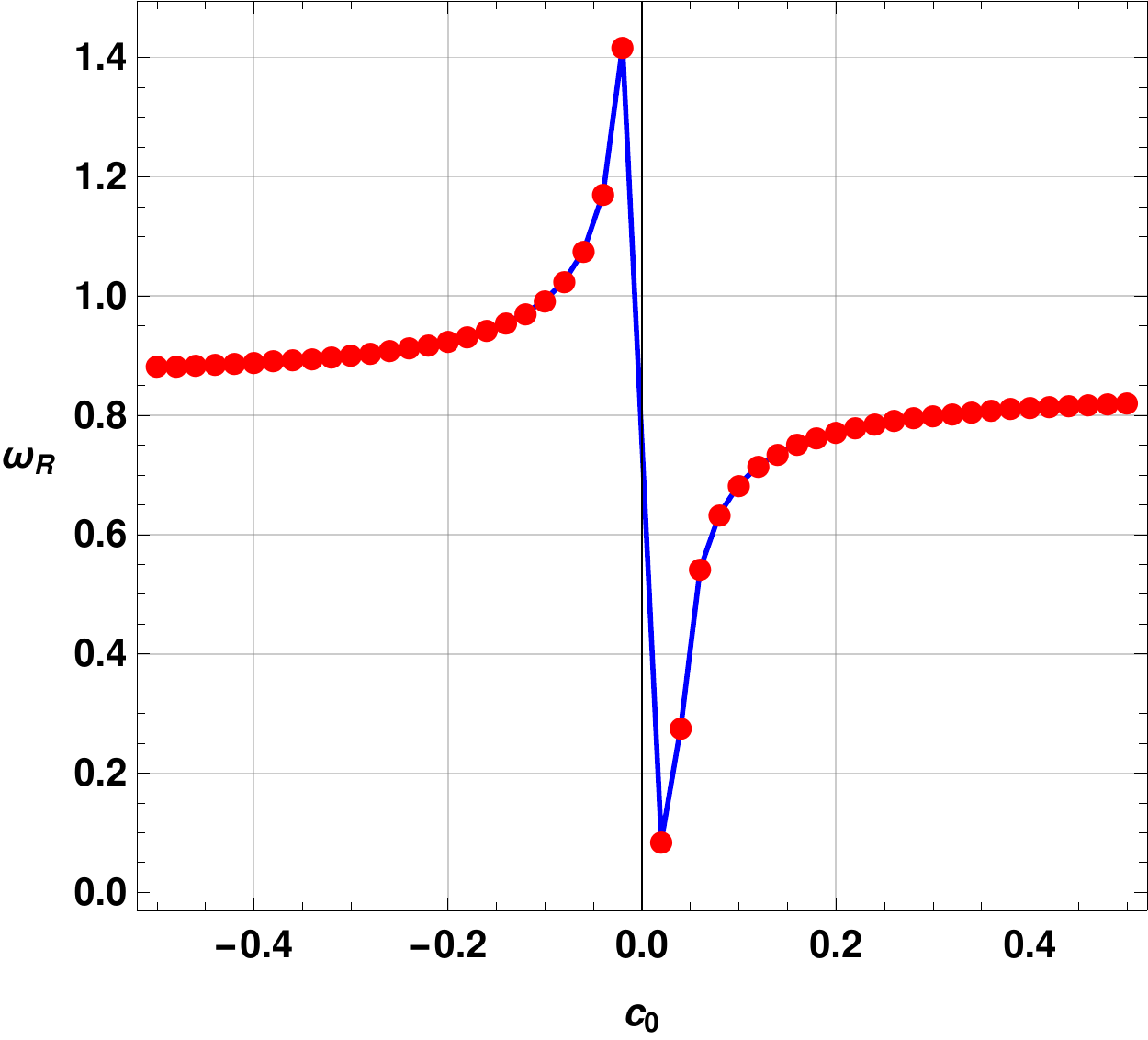}\hspace{0.5cm}
   \includegraphics[scale = 0.5]{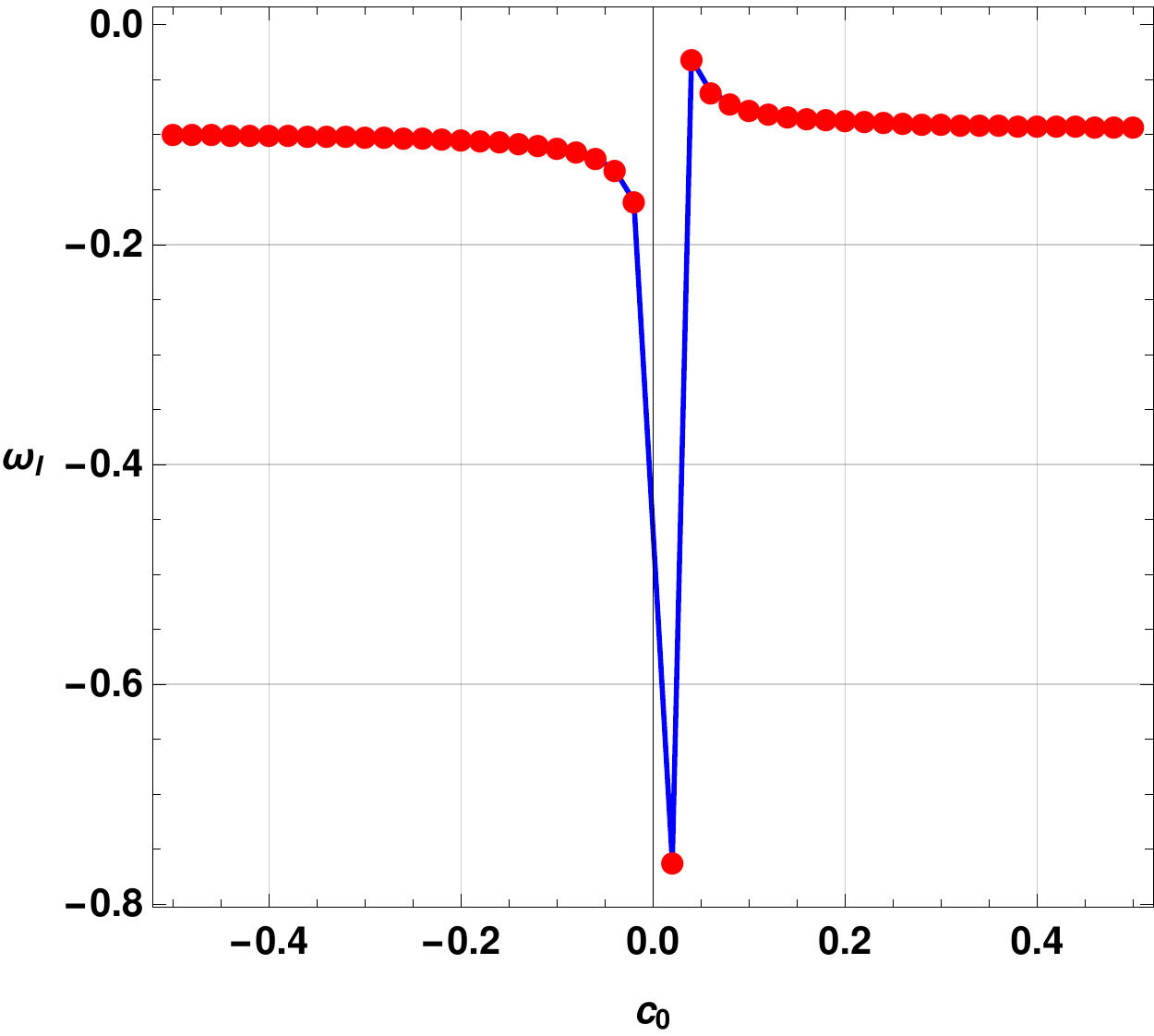}} \vspace{-0.2cm}
\caption{The real (left panel) and imaginary (right panel) parts of the QNMs of the symmergent black hole for electromagnetic perturbations as a function of the symmergent parameter $c_{\rm O}$ with $M=1$, $G = 1$, $n= 0$, $l=4$ and $\alpha = 0.9$.}
\label{QNMs04}
\end{figure}

To see the impacts of the model parameters on the QNM spectrum, we have explicitly plotted the real and imaginary QNMs with respect to the model parameters. For this purpose, we have utilised the Pad\'e averaged 6th-order WKB approximation method with a higher value of $l$. The reason behind choosing a comparatively higher value of multipole moment $l$ is that the error associated with the WKB method for higher $l$ is significantly less.

In Fig. \ref{QNMs01}, we have shown the variation of real (on left) and imaginary (on right) QNMs with respect to the model parameter $\alpha$ for scalar perturbation. It is seen that with an increase in the value of $\alpha$, real QNMs or oscillation frequencies increase significantly. The variation is almost linear. One may note that $\alpha<1$ implies de Sitter spacetime. Thus, a de Sitter black hole shows smaller oscillation frequencies than those obtained in an anti-de Sitter or asymptotically flat scenario. Similarly, the damping rate or decay rate of GWs also increases significantly with an increase in the value of $\alpha$. The variation of the damping rate also follows an almost linear pattern with respect to $\alpha$.

In Fig. \ref{QNMs02}, we have shown the variation of ringdown GWs frequency and damping rate with respect to the model parameter $c_{\rm O}$ for scalar perturbation. For positive values of $c_{\rm O}$, both real quasinormal frequencies and damping rate increase non-linearly. The variation is more significant for smaller values of $c_{\rm O}$ and for large values near $0.4$, the variation of both real and imaginary QNMs is very small. In the case of negative values of $c_{\rm O}$, both damping rate and oscillation frequency increase non-linearly and as $c_{\rm O}$ approaches $0$, both values increase to a maximum. It is worth mentioning that at $c_{\rm O} = 0$, we observe a discontinuity in the QNM spectrum.

Similarly, in Fig. \ref{QNMs03} and \ref{QNMs04}, we have shown the corresponding graphs for electromagnetic perturbation. As predicted from the potential behaviour, we see a similar variation pattern of electromagnetic QNMs with respect to the model parameters as we obtained for scalar perturbation. However, in the case of electromagnetic perturbation, the oscillation frequencies and damping rates are comparatively smaller in comparison to those obtained in the case of scalar perturbation. 

{\color{black} Our investigation shows that both parameters have distinct and different impacts on the QNMs. Although the black hole metric can effectively  behave as a dS or AdS black hole, one  notes that the impacts of the symmergent parameters on the QNM spectrum are different. The signature of symmergent parameters on QNMs is noticeably different from other black hole hairs in different well-studied black hole systems, including the standard Schwarzschild dS/AdS black holes \cite{Yang:2022ifo,Daghigh:2008jz,Zhidenko:2003wq,Zhidenko:2005mv}. It is already mentioned in the previous part of this paper that symmergent gravity mimics $f(R)$ gravity of type $R+R^2$ with non-zero cosmological constant. However, the variations of QNMs with model parameters found in this study are not similar to those studied in similar aspects for black hole and wormhole configurations \cite{Ovgun188, gogoi_wormhole}. These findings emphasize the uniqueness of the symmergent parameters and their discernible influence on the QNM behaviour. }

\section{Evolution of Scalar and Electromagnetic Perturbations on the Black
hole geometry}\label{sec6} 

In the preceding section, we have performed numerical computations to determine the characteristics of the QNMs and examined their dependence on the model parameters $\alpha$ and $c_{\rm O}$. In the subsequent section, our focus shifts towards the temporal profiles of the scalar perturbation and electromagnetic perturbation. To obtain these profiles, we employ the time domain integration framework as outlined in the work by Gundlach et al. \cite{gundlach}. To proceed further, we define $\psi(r_*,
t) = \psi(i \Delta r_*, j \Delta t) = \psi_{i,j} $ and $V(r(r_*)) = V(r_*,t) =
V_{i,j}$. After that we express Eq. \eqref{scalar_KG} as
\begin{equation}
\dfrac{\psi_{i+1,j} - 2\psi_{i,j} + \psi_{i-1,j}}{\Delta r_*^2} - \dfrac{%
\psi_{i,j+1} - 2\psi_{i,j} + \psi_{i,j-1}}{\Delta t^2} - V_i\psi_{i,j} = 0.
\end{equation}
The initial conditions are $\psi(r_*,t) = \exp \left[ -\dfrac{(r_*-k_1)^2%
}{2\sigma^2} \right]$ and $\psi(r_*,t)\vert_{t<0} = 0$ (here $k_1$ and $%
\sigma$ are median and width of the initial wave-packet). Using these we calculate the time evolution associated with the scalar perturbation as 
\begin{equation}
\psi_{i,j+1} = -\,\psi_{i, j-1} + \left( \dfrac{\Delta t}{\Delta r_*}
\right)^2 (\psi_{i+1, j + \psi_{i-1, j}}) + \left( 2-2\left( \dfrac{\Delta t%
}{\Delta r_*} \right)^2 - V_i \Delta t^2 \right) \psi_{i,j}.
\end{equation}
Implementing this iteration scheme with a fixed value of $\frac{\Delta t}{\Delta r_*}$, it is possible to obtain the time domain profiles numerically. But one needs to keep $\frac{\Delta t}{\Delta r_*}
< 1$ in order to satisfy the Von Neumann stability condition during the numerical procedure.

\begin{figure}[htbp]
\centerline{
   \includegraphics[scale = 0.8]{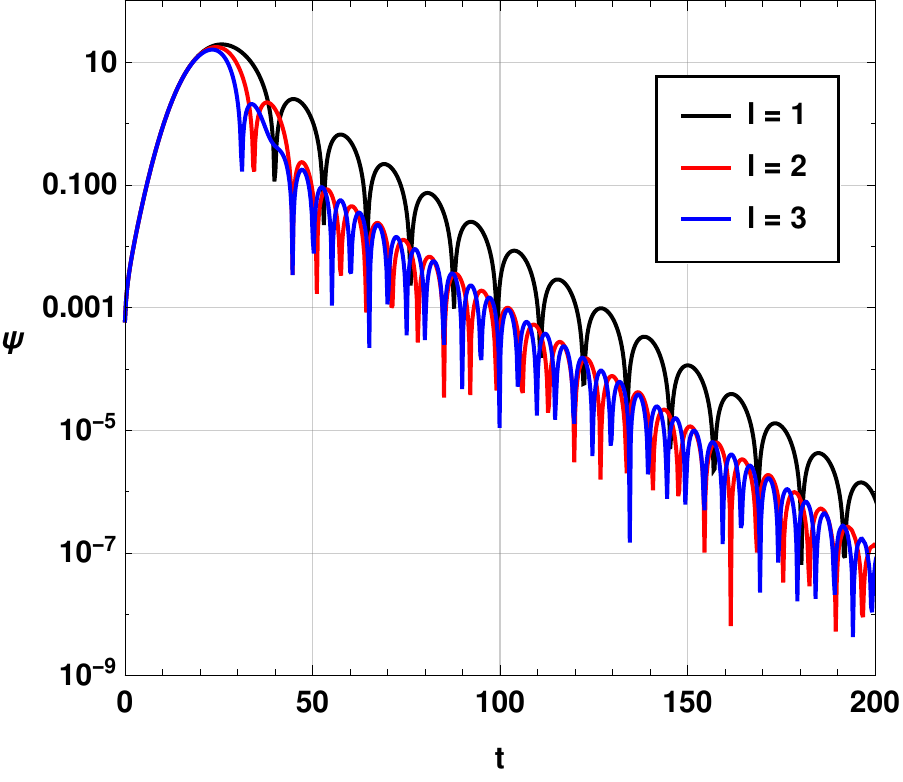}\hspace{0.5cm}
   \includegraphics[scale = 0.8]{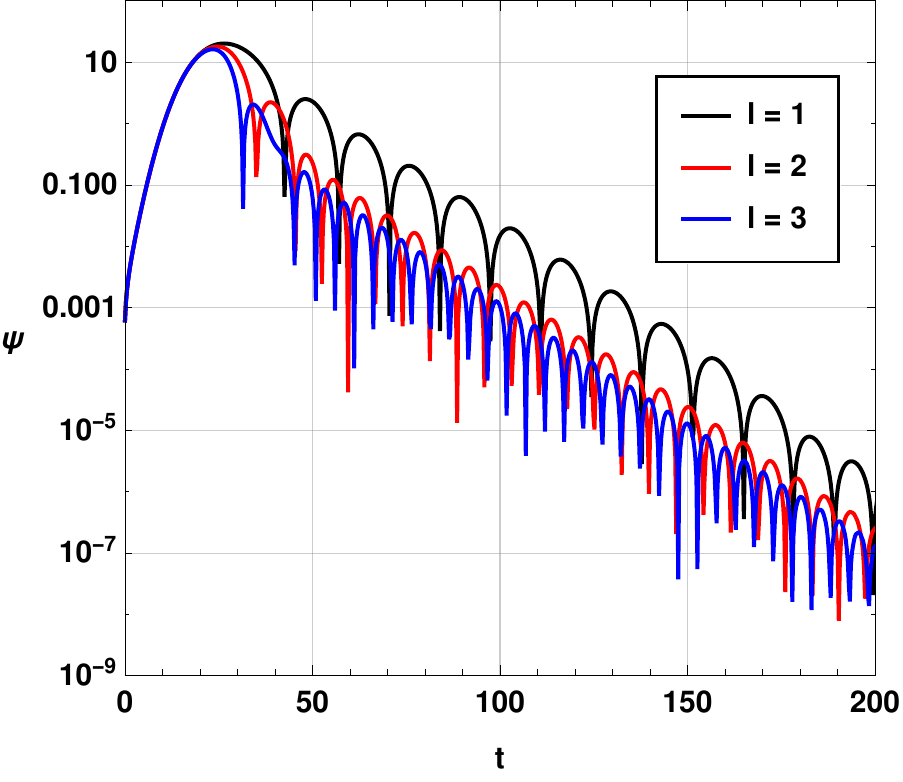}} \vspace{-0.2cm}
\caption{The time-domain profiles of the massless scalar
perturbations (first panel) and electromagnetic perturbations (right panel) for different multipole moments $l$ with the parameter values $M=1$, $G = 1$, $n= 0, \alpha = 0.9, c_{\rm O}= 0.3$.}
\label{time01}
\end{figure}

\begin{figure}[htbp]
\centerline{
   \includegraphics[scale = 0.8]{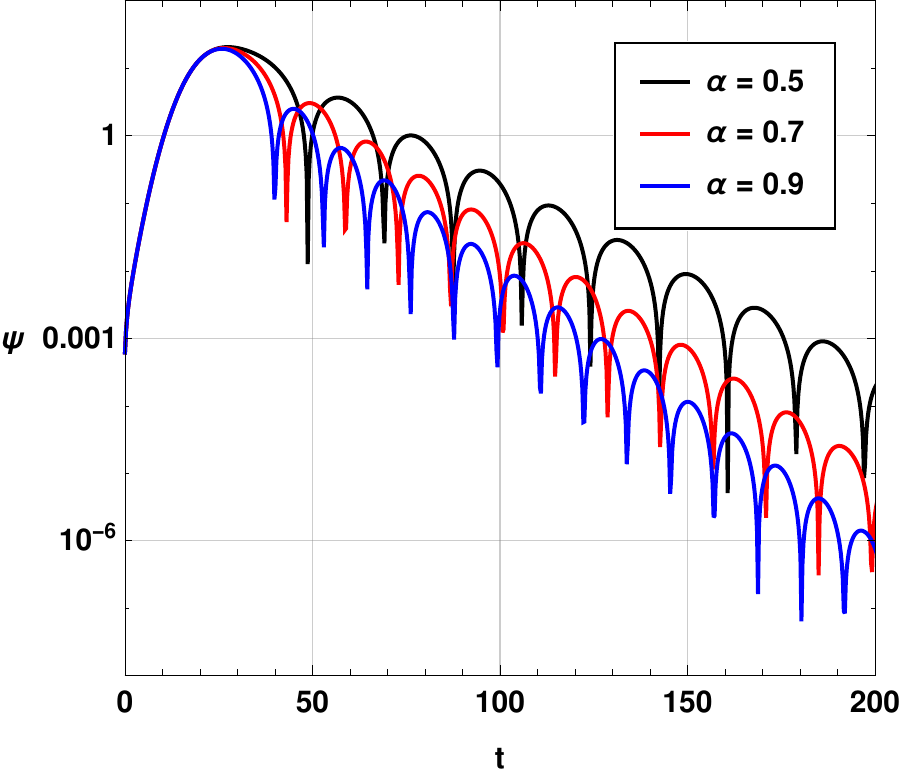}\hspace{0.5cm}
   \includegraphics[scale = 0.8]{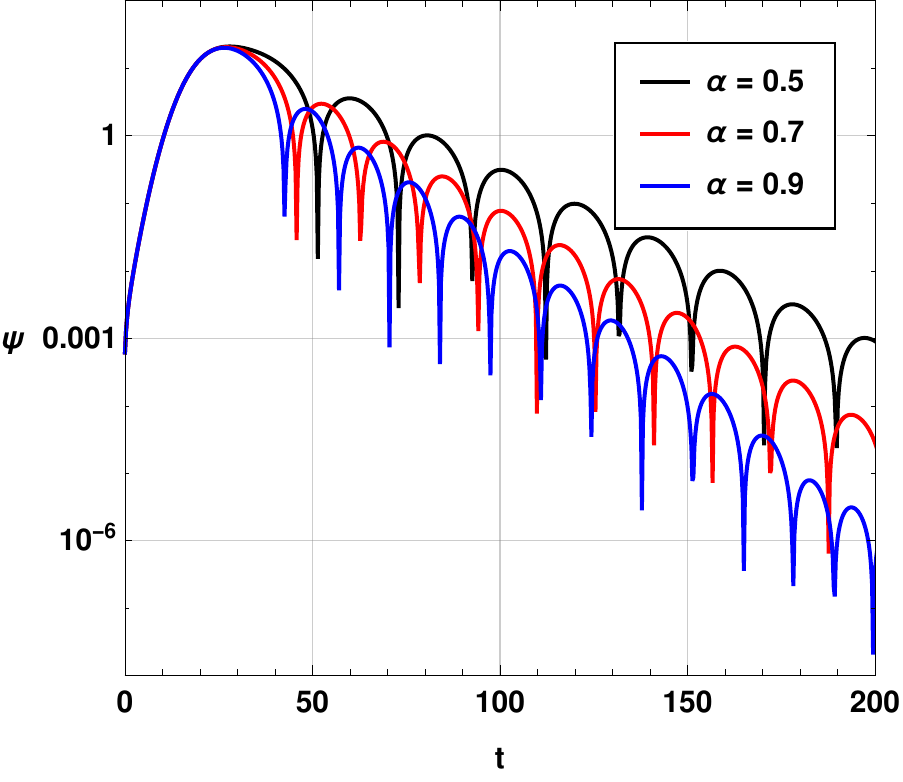}} \vspace{-0.2cm}
\caption{The time-domain profiles of the massless scalar
perturbations (first panel) and electromagnetic perturbations (right panel) for different vacuum energy parameters $\alpha$ with the parameter values $M=1$, $G = 1$, $l=1$, $n= 0,  c_{\rm O}= 0.3$.}
\label{time02}
\end{figure}

\begin{figure}[htbp]
\centerline{
   \includegraphics[scale = 0.8]{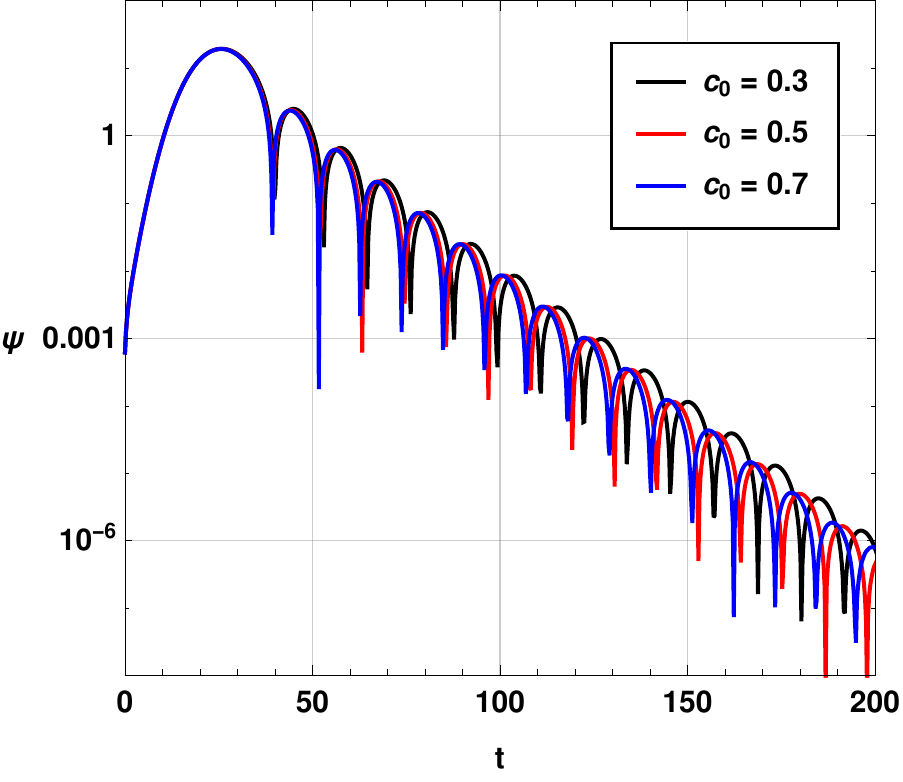}\hspace{0.5cm}
   \includegraphics[scale = 0.8]{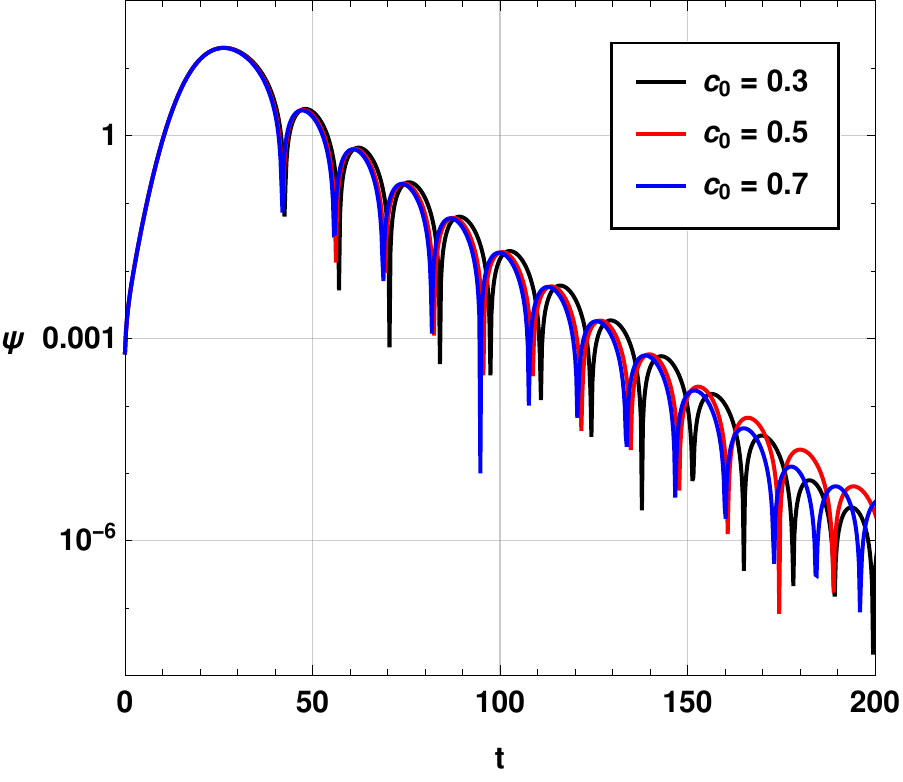}} \vspace{-0.2cm}
\caption{The time-domain profiles of the massless scalar
perturbations (first panel) and electromagnetic perturbations (right panel) for different symmergent loop factors $c_{\rm O}$ with the parameter values $M=1$, $G = 1$, $l=1$, $n= 0, \alpha = 0.9$.}
\label{time03}
\end{figure}

In the left panel of Fig. \ref{time01}, we have depicted the temporal representation of the scalar perturbation, while in the right panel, we have illustrated the temporal profile of the electromagnetic perturbation. The chosen parameters include an overtone number of $n=0$, $\alpha=0.9$, and $c_{\rm O}=0.3$, with varying values of the multipole moment $l$. Notably, the time domain profiles exhibit significant distinctions across different values of $l$. In both cases, an increase in $l$ corresponds to a higher frequency. However, the decay rate of the scalar perturbation appears to escalate as $l$ increases, whereas the variation is marginal for the electromagnetic perturbation. It is noteworthy that the damping or decay rate for the scalar perturbation is higher compared to that of the electromagnetic perturbation.

In Fig. \ref{time02}, we have graphically presented the temporal profiles for both the scalar perturbation (left panel) and the electromagnetic perturbation (right panel) while varying the parameter $\alpha$. These profiles were obtained with an overtone number of $n=0$, a multipole number of $l=1$, and $c_{\rm O}=0.3$. It is evident that the decay rate of the scalar perturbation's time profile is greater in comparison to that of the electromagnetic perturbation. These findings align consistently with the earlier observations made regarding the QNMs.

Finally, in Fig. \ref{time03}, we have shown the time domain profiles for both scalar (on left) and electromagnetic (on right) perturbations with different values of the model parameter $c_{\rm O}$. One may note that for smaller values of time $t$, the differences in the time profile are very small for both types of perturbations. The variation in the oscillation frequencies is observable for higher values of $t$ in the time profile. However, the damping rate seems to be less affected by the parameter $c_{\rm O}$ for the considered values of the parameter. This stands in agreement with the variation of QNM plots studied in the previous part of the paper.

\section{Greybody factors using the WKB approach}

In the scientific literature, it has been established by Hawking (1975) \cite{Hawking:1975vcx} that black holes emit radiation, known as Hawking radiation, which renders them non-black and gives them a grey appearance  \cite{Singleton:2011vh,Akhmedova:2008dz}. The observation of this radiation by an asymptotic observer differs from the original radiation near the black hole's horizon due to a redshift factor, referred to as the greybody factor. Several methods have been employed to calculate greybody factors, including works by Maldacena et al. (1996) \cite{Maldacena:1996ix}, Cvetic et al. (1997) \cite{Cvetic:1997uw}, Fernando (2004) \cite{Fernando:2004ay}, and others \cite{Okyay2022,Ovgun188,Pantig:2022gih,Yang:2022xxh,Yang:2022ifo,Panotopoulos:2018pvu,Panotopoulos:2016wuu,Rincon:2018ktz,Ahmed:2016lou,Javed:2021ymu,Javed:2022kzf,Mangut:2023oou,Al-Badawi:2022aby}. In this part, we intend to utilise higher order WKB method to calculate the greybody factors for both scalar and electromagnetic perturbations.

We want to investigate the wave equation (\ref{radial_scalar}) while considering boundary conditions that permit the presence of incoming waves originating from infinity. Due to the symmetrical nature of the scattering characteristics, this scenario is equivalent to studying the scattering of a wave approaching from the horizon. This equivalence is appropriate when determining the proportion of particles that are reflected back from the potential barrier towards the horizon. The scattering boundary conditions for equation (\ref{radial_scalar}) are expressed as follows:
\begin{equation}\label{BC}
\begin{array}{ccll}
    \Psi &=& e^{-i\omega r_*} + R e^{i\omega r_*},& r_* \rightarrow +\infty, \\
    \Psi &=& T e^{-i\omega r_*},& r_* \rightarrow -\infty, \\
\end{array}%
\end{equation}
where $R$ and $T$ are the reflection and transmission coefficients, satisfying
\begin{equation}\label{1}
\left|T\right|^2 + \left|R\right|^2 = 1
\end{equation}
on the basis of the probability conservation.
Once we determine the reflection coefficient, we can calculate the transmission coefficient for each multipole number $l$ using the WKB approach. In fact, the transmission coefficient is defined to be the greybody factor $A$ of the black hole:
\begin{equation}
\left|A\right|^2=1-\left|R\right|^2=\left|T\right|^2.
\end{equation}
where the WKB approach gives
\begin{equation}\label{moderate-omega-wkb}
R = (1 + e^{- 2 i \pi K})^{-\frac{1}{2}},
\end{equation}
such the phase factor $K$ is determined from the following equation:
\begin{equation}
K - i \frac{(\omega^2 - V_{0})}{\sqrt{-2 V_{0}^{\prime \prime}}} - \sum_{i=2}^{i=6} \Lambda_{i}(K) =0.
\end{equation}
In the above equation, $V_0$ represents the maximum value of the effective potential, $V_{0}^{\prime \prime}$ is the second derivative of the effective potential at its maximum with respect to the tortoise coordinate $r_{*}$, and $\Lambda_i$ represents higher-order WKB corrections that depend on up to the $2i$th order derivatives of the effective potential at its maximum. These corrections have been discussed in works such as \cite{Schutz,Will_wkb,Konoplya_wkb,Maty_wkb}, and they depend on $K$. The 6th order WKB formula, as presented in \cite{Konoplya_wkb}, is primarily used in this study, although lower orders may be applied for lower frequencies and multipole numbers. It is worth noting that the accuracy of the WKB approach is compromised at small frequencies where reflection is nearly total and the grey-body factors approach zero. However, this inaccuracy does not significantly impact the estimation of energy emission rates.

As the WKB method is well-established and widely known (for more detailed information, refer to reviews such as \cite{Konoplya:2019hlu,Konoplya:2011qq} and the references therein), we will not provide a comprehensive description of it here.

\begin{figure}[htbp]
\centerline{
   \includegraphics[scale = 0.8]{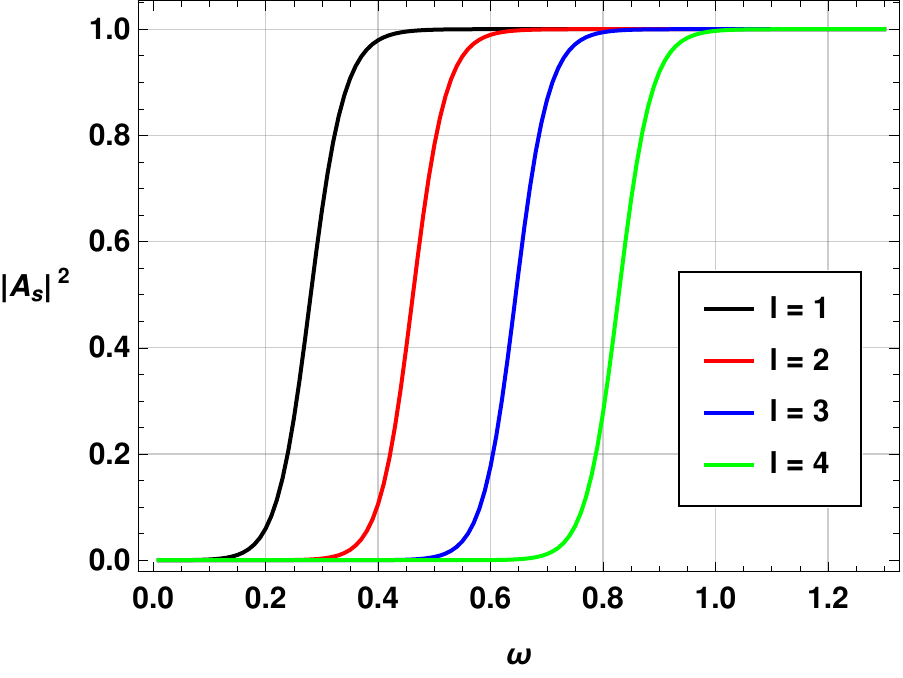}\hspace{0.5cm}
   \includegraphics[scale = 0.8]{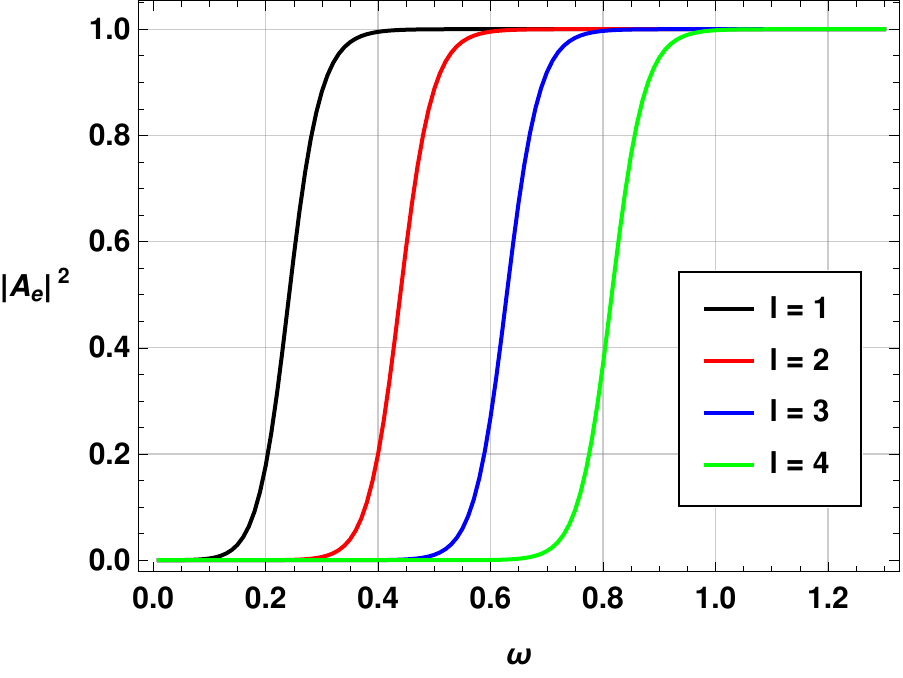}} \vspace{-0.2cm}
\caption{The greybody factors for massless scalar (left panel) and electromagnetic (right panel) perturbations for different values of the multipole moment $l$ with the parameter values $M=1$, $G = 1$, $\alpha = 0.9$ and $c_{\rm O} = 0.4$.}
\label{G01}
\end{figure}

\begin{figure}[htbp]
\centerline{
   \includegraphics[scale = 0.8]{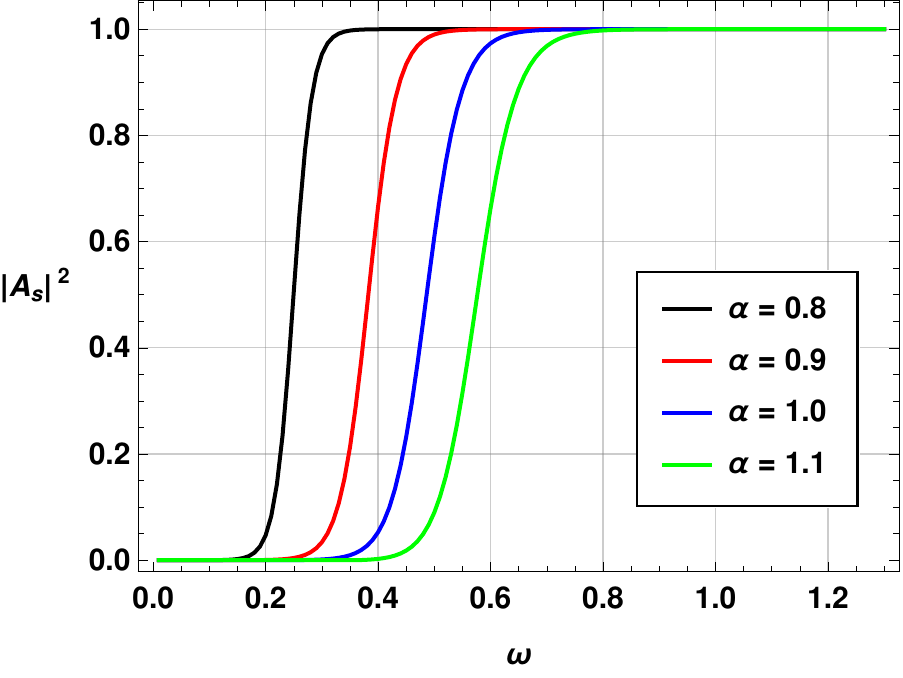}\hspace{0.5cm}
   \includegraphics[scale = 0.8]{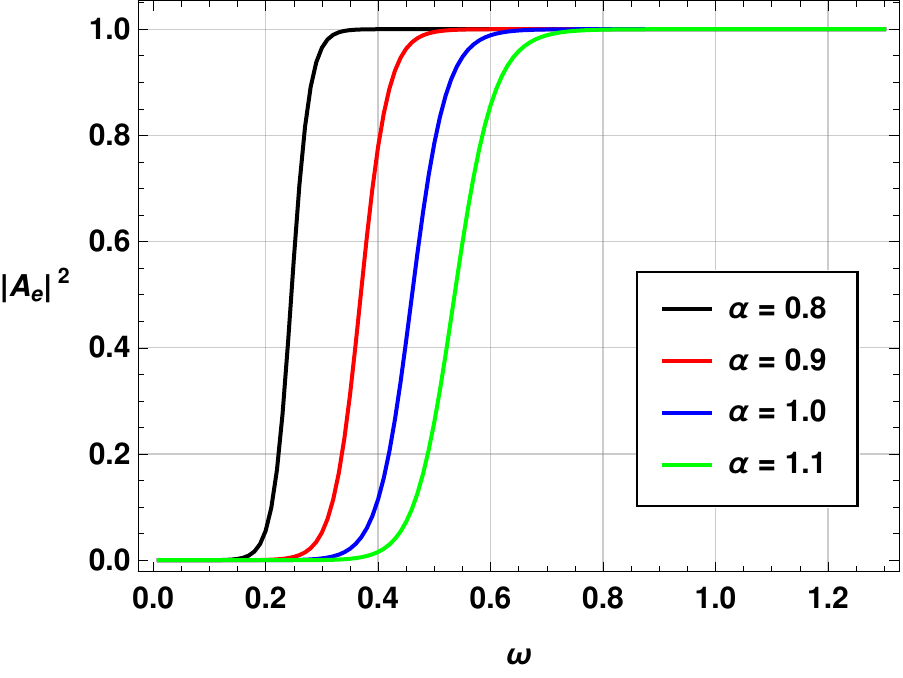}} \vspace{-0.2cm}
\caption{The greybody factors for massless scalar (left panel) and electromagnetic (right panel) perturbations for different values of the vacuum energy parameter $\alpha$ with the parameter values $M=1$, $G = 1$, $l = 2$ and $c_{\rm O} = 0.1$.}
\label{G02}
\end{figure}

\begin{figure}[htbp]
\centerline{
   \includegraphics[scale = 0.8]{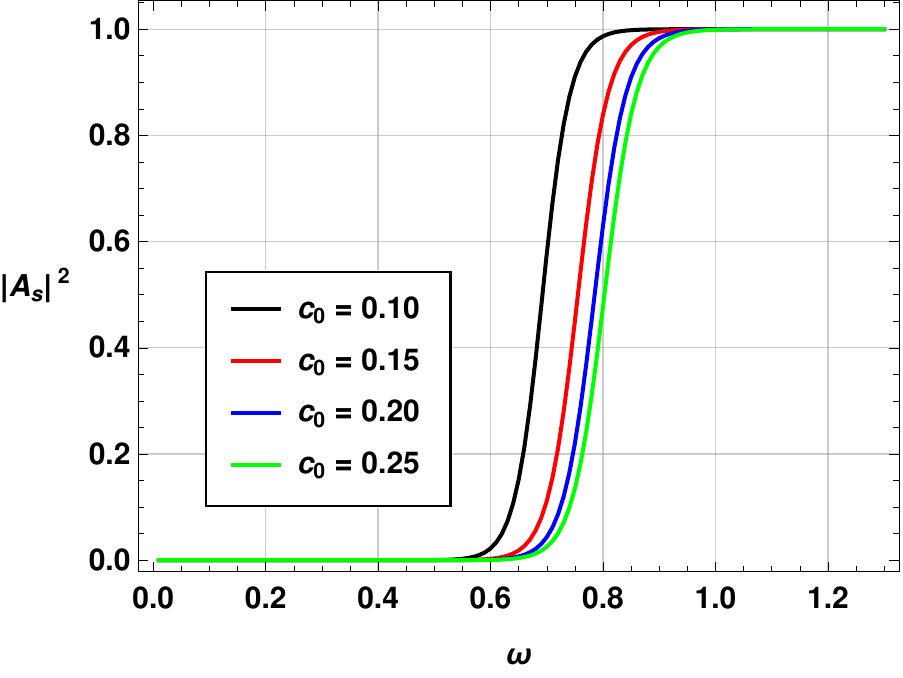}\hspace{0.5cm}
   \includegraphics[scale = 0.8]{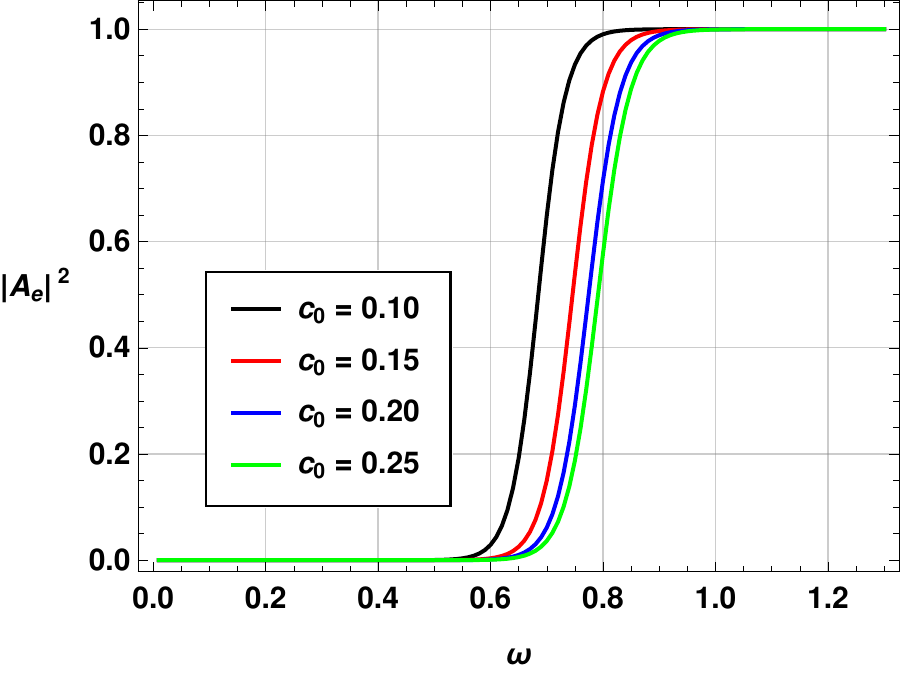}} \vspace{-0.2cm}
\caption{The greybody factors for massless scalar (left panel) and electromagnetic (right panel) perturbations for different values of the symmergent parameter $c_{\rm O}$ with the parameter values $M=1$, $G = 1$, $\alpha = 0.9$ and $l = 4$.}
\label{G03}
\end{figure}

In Figs. \ref{G01}, \ref{G02}, and \ref{G03} we have plotted the greybody factors for both scalar and electromagnetic perturbations. In these Figs., $A_s$ represents greybody factors for scalar perturbation while $A_e$ represents greybody factors for electromagnetic perturbation. From Fig. \ref{G01}, we observe that the variation of greybody factors is almost identical for both types of perturbations for different values of the multipole moments $l$. A close observation reveals that in the case of electromagnetic perturbations, greybody factors attain a maximum for comparatively smaller values of $\omega$.

In Fig. \ref{G02}, we show the impacts of model parameter $\alpha$ on the greybody factors. The figure reveals that the greybody factors are higher for smaller values of $\alpha$ for both types of perturbations. 
{\color{black} The greybody factors represent the probability that a particle or wave is absorbed or scattered by a black hole. In this context, the model parameter $\alpha$ plays a crucial role in determining the behaviour of the greybody factors. The observation that the greybody factors are higher for smaller values of $\alpha$ indicates that for certain types of perturbations, black holes with smaller $\alpha$ values exhibit stronger absorption and scattering of particles or waves. This suggests that when the symmergent parameter $\alpha$ is reduced, the black hole becomes more effective at capturing and interacting with incoming matter or radiation.}

A similar observation is done for the parameter $c_{\rm O}$ in Fig. \ref{G03}. 
{\color{black} This sensitivity suggests that near smaller values of $c_{\rm O}$, the black hole's ability to absorb and scatter particles or waves is highly responsive to slight changes in $c_{\rm O}$, resulting in significant variations in the greybody factors. On the other hand, as $c_{\rm O}$ increases, the impact of this parameter on the greybody factors diminishes. Consequently, for larger values of $c_{\rm O}$, the black hole's behaviour regarding absorption and scattering becomes less sensitive to changes in $c_{\rm O}$.}

{\color{black} In essence, these findings underscore the significance of model parameters $\alpha$ and $c_{\rm O}$ in shaping the intricate interplay between black holes and external particles or waves. Gaining insight into these influences can facilitate a deeper grasp of the fundamental characteristics of black holes in the context of symmergent gravity, elucidating their responses under various perturbations. By shedding light on such behaviour, these investigations contribute to a more comprehensive understanding of the intriguing dynamics surrounding black holes in symmergent gravity.}

\subsection{Rigorous Bounds on Greybody Factors} \label{sec5}

In this part of our investigation, we consider rigorous bounds on greybody factors by utilising a different method. Since, this part of our investigation reveals similar behaviour of both scalar and electromagnetic perturbations in terms of greybody factors, in the next part we shall take into account scalar perturbation only.

Visser (1998)  \cite{Visser:1998ke}and subsequently Boonserm and Visser (2008) \cite{Boonserm:2008zg} presented an elegant analytical approach to obtain rigorous bounds on greybody factors. These bounds have been further investigated by Boonserm et al. (2017, 2019) \cite{Boonserm:2017qcq},  Yang et al. (2022) \cite{Yang:2022ifo}, Gray et al. (2015) \cite{Gray:2015xig}, Ngampitipan et al. (2012) \cite{Ngampitipan:2012dq}, and others \cite{Chowdhury:2020bdi,Miao:2017jtr,Liu:2021xfs,Barman:2019vst,Xu:2019krv,Boonserm:2017qcq}.

In this study, our focus is specifically on the bounds for the greybody factors of symmergent black holes. To achieve this, we analyze the Klein-Gordon equation for the massless scalar field, similarly discussed in Section III A, and then examine the reduced effective potential denoted as $V_{s}(r)$ in Equation \ref{Vs} reduced to
\begin{equation}
V(r) = \frac{l(l + 1)f(r)}{r^{2}} + \frac{f(r)f'(r)}{r}.\label{poten}
\end{equation}

Subsequently, utilizing the aforementioned effective potential, we investigate the lower rigorous bound of the graybody factor for symmergent black holes, aiming to examine the impact of $\epsilon$ on the bound. The formula for deriving the rigorous bound of the greybody factor is provided below \cite{Visser:1998ke,Boonserm:2008zg}:
\begin{equation} \label{bound}
A_{b} \geq \operatorname{sech}^{2}\left(\frac{1}{2 \omega} \int_{-\infty}^{\infty}\left|V\right| \frac{d r}{f(r)} \right)
\end{equation}
where we keep in mind that $A_b=T_b$, $T$ being the transmission coefficient.

Moreover, the boundary conditions for the aforementioned formula are modified to account for the presence of the cosmological constant, as outlined in the work by Boonserm et al. (2019) \cite{Boonserm:2019mon}. The modified boundary conditions are given as follows:
\begin{equation}
A \geq A_{b}=\operatorname{sech}^{2}\left(\frac{1}{2 \omega} \int_{r_{H}}^{R_{H}} \frac{|V|}{f(r)} d r\right)=\operatorname{sech}^{2}\left(\frac{A_{l}}{2 \omega}\right),
\end{equation}
with \begin{equation}
A_{l}=\int_{r_{H}}^{R_{H}} \frac{|V|}{f(r)} d r=\int_{r_{H}}^{R_{H}}\left|\frac{l(l+1)}{r^{2}}+\frac{f^{\prime}}{r}\right| d r.
\end{equation}

Therefore, we have successfully computed the rigorous bounds on greybody factors for symmergent black hole
\begin{equation}
A_{b}=\text{sech}^2\left(\frac{\left(r_{\text{H}}-R_H\right) \left(12 \pi  c G^2 M R_H+12 \pi  c G r_{\text{H}} \left(G M+l (l+1) R_H\right)+(\alpha -1) R_H^2 r_{\text{H}}^2\right)}{24 \pi  c G \omega  R_H^2 r_{\text{H}}^2}\right)
\end{equation}
as a function of various parameters, including $c_{\rm O}$, $\alpha$, and others.

\begin{figure}[htbp]
\centerline{
   \includegraphics[scale = 0.8]{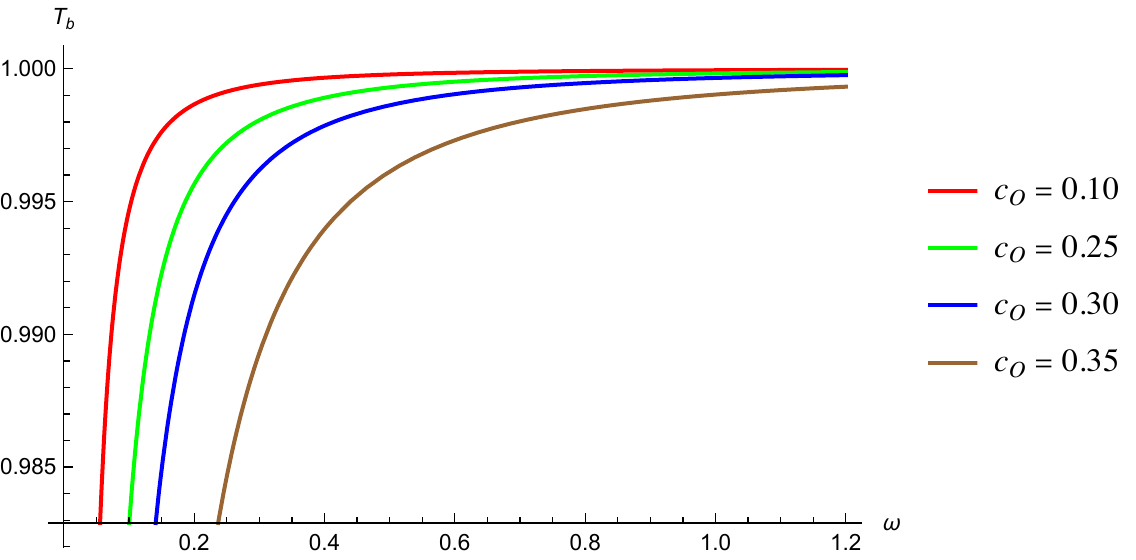}\hspace{0.5cm}
   \includegraphics[scale = 0.8]{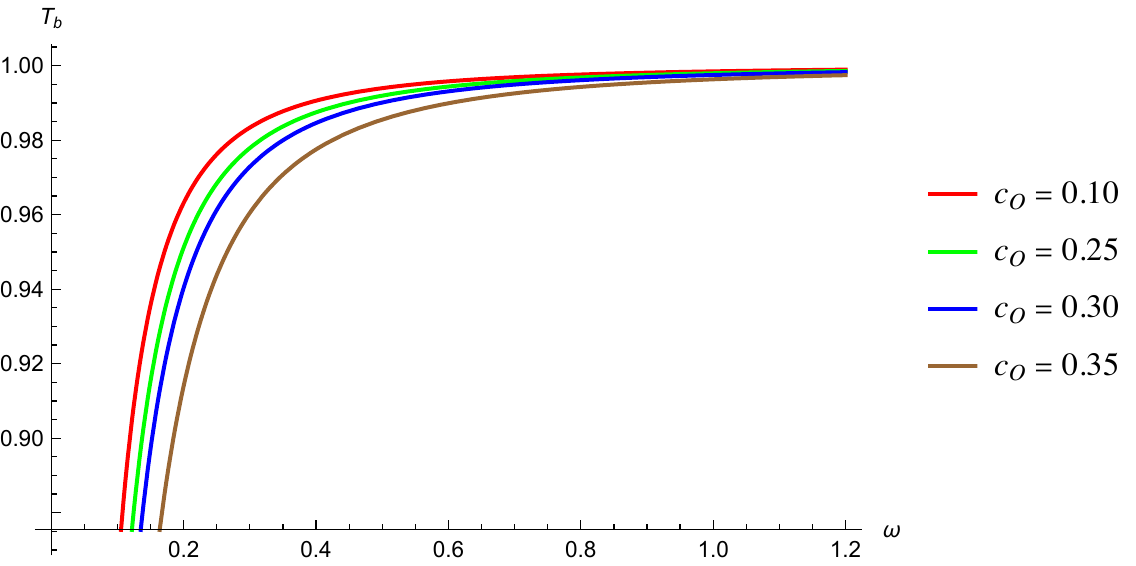}} \vspace{-0.2cm}
   
\caption{The greybody bound $A_b$ as a function of the frequency for different values of the symmergent parameter $c_{\rm O}$ for $\alpha=0.5$ (left panel) and $\alpha=0.8$ (right panel).}
\label{G04}
\end{figure}

\begin{figure}[htbp]
\centerline{
   \includegraphics[scale = 0.8]{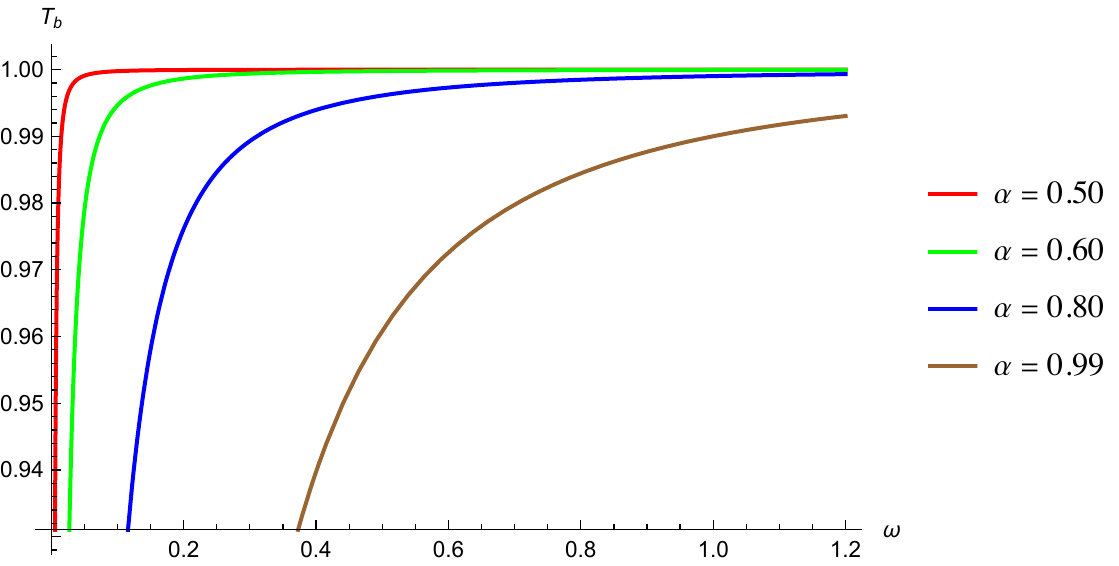}\hspace{0.5cm}
   \includegraphics[scale = 0.8]{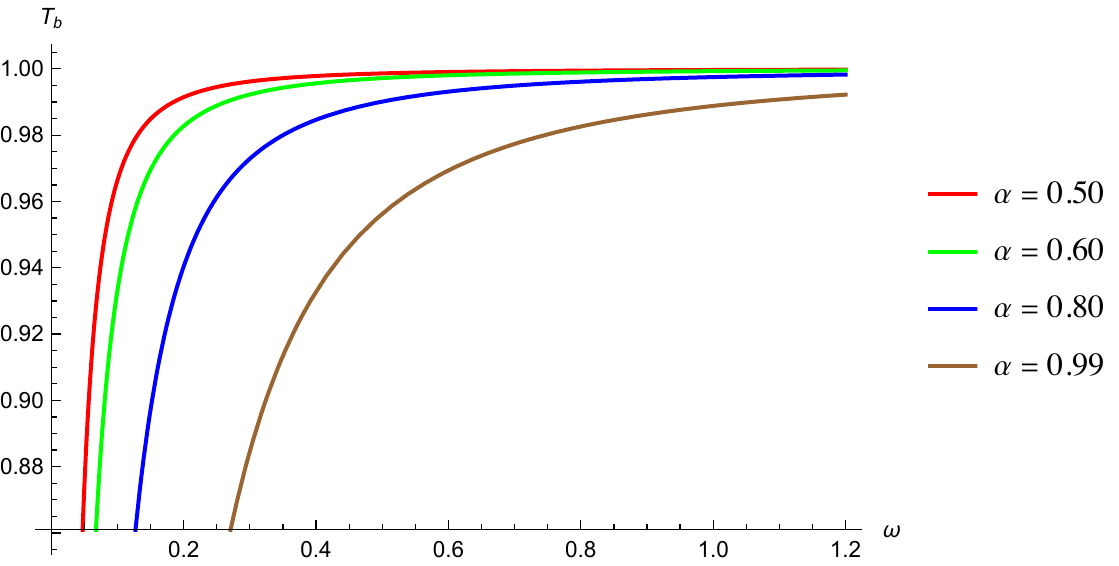}} \vspace{-0.2cm}
   
\caption{The greybody bound $A_b$ as a function of the frequency for different values of the vacuum energy parameter $\alpha$ for $c_{\rm O}=0.2$ (left panel) and $c_{\rm O}=0.35$ (right panel).}
\label{G05}
\end{figure}

By performing numerical calculations, we can evaluate the bound and visualize it in Figure \ref{G04} for the case of $l=0$ and $c_o=0.2$, and in Figure \ref{G05} for the case of $l=0$ and $c_o=0.35$. The resulting graphs indicate that as the parameter $c_o$ increases, the bound on the greybody factor decreases. This observation suggests that symmergent black holes exhibit stronger barrier properties and possess lower greybody factors compared to Schwarzschild black holes.

\section{Conclusion}\label{sec7} 

In this study, we have studied black hole solutions in symmergent gravity in regard to their QNMs and greybody factors. Our analyses reveal that the quadratic curvature term $c_{\rm O}$ has significant impacts on the QNMs. Indeed, real QNMs get smaller and tend to zero (become larger)  when the symmergent parameter $c_{\rm O}$ takes small positive (small negative) values. One notes that for asymptotic values of $c_{\rm O}$ (large values towards +$\infty$ as well as small values towards $- \infty$) QNMs attain constant values corresponding to their values in the Schwarzschild black hole. The vacuum energy parameter $\alpha$ (specific to the symmergence) has noticeable impact on the QNM spectrum in that both the real QNMs and the damping rates vary almost linearly with the parameter $\alpha$. Also, real QNMs and the damping rates in the de Sitter case are both smaller than those in the anti-de Sitter case. 

Another result obtained from this investigation is that the  QNMs obtained by the AIM and Pad\'e averaged 6th order WKB method exhibit small variations (good agreement) at small (large) values of multipole moment $l$. The percentage deviations of QNMs obtained by AIM and WKB method is relatively large for the electromagnetic perturbation (around $0.0114\%$ for $l=1$) compared to the massless scalar perturbation. The time domain profiles of both types of perturbations are in agreement with the previous results.

We have used the WKB formula to compute the greybody factors. It turns out that greybody factors can't be used to probe the scalar and vector perturbations since they remain similar for both types of perturbations.  We found that the symmergent parameter $\alpha$ can have noticeable impact on the greybody factors. Indeed, greybody factor (tramsmission coefficient) gets larger as $\alpha$ decreases. Similar behavior is noticed also for the quadratic curvature coefficient $c_{\rm O}$. Finally, we computed rigorous bounds on the greybody factors for scalar perturbations. The dependencies of the bounds on model parameters remain similar to those of the greybody factors.

 {\color{black} One may note that ground-based GW detectors like LIGO may not be able to detect QNMs with suitable precision \cite{Ferrari2008}. In Ref.s \cite{Ferrari2008, gogoi_bumblebee} it is shown that space-based GWs detectors like LISA can be sensitive enough to detect QNMs from different prominent sources.}
Hence we believe that in the near future, with the help of LISA, QNMs can be detected more precisely and those observational results can be used to constrain symmergent gravity further.  {\color{black} Our study, along with the observational results of QNMs can help us to understand symmergent gravity in more detail. Observational constraints on QNMs in symmergent gravity from LISA along with shadow constraints from Event Horizon Telescope (EHT) can be used in the near future to check the consistencies and feasibility of the theory.}

\section{ACKNOWLEDGMENTS}

A. {\"O}. would like to acknowledge the contribution of the COST Action
CA18108 - Quantum gravity phenomenology in the multi-messenger approach
(QG-MM).  A. {\"O}. and D. D.  would like to acknowledge networking support by the COST Action CA21106 - COSMIC WISPers in the Dark Universe: Theory, astrophysics and experiments (CosmicWISPers). DJG would like to thank Prof. U. D. Goswami for some useful discussions.

%%%%%%%%%%%%%%%%%%%%%%%%%%%
%%  PLEASE KEEP EACH BIBITEM IN A SINGLE LINE WITHOUT USING 'ENTER'
%%%%%%%%%%%%%%%%%%%%%%%%%%%

\end{document}